\documentclass{jfm}

\usepackage{amsmath}
\usepackage{amsfonts}
\usepackage{bm}
\usepackage{enumerate}
\usepackage{graphicx}
\usepackage{mathrsfs}
\usepackage{subfigure}
\usepackage{url}
\usepackage{verbatim}
\usepackage{amssymb}
\usepackage[pdftex,bookmarksnumbered]{hyperref}
\usepackage{color}

\usepackage{epstopdf, epsfig}

\shorttitle{Shear dispersion of multispecies electrolyte solutions in the channel domain}
\shortauthor{Lingyun Ding}

\title{Shear dispersion of multispecies electrolyte solutions in channel domain}

\author{Lingyun Ding\aff{1}
  \corresp{\email{dingly@g.ucla.edu}}}

  \affiliation{\aff{1}Department of Mathematics, University of California Los Angeles, CA, 90095, United States}

\begin{document}
\maketitle

\begin{abstract}
  In multispecies electrolyte solutions, even in the absence of an external electric field, differences in ion diffusivities induce an electric potential and generate additional fluxes for each species. This electro-diffusion process is well-described by the advection-Nernst-Planck equation. This study aims to analyze the long-time behavior of the governing equation under electroneutrality and zero current conditions and investigate how the diffusion-induced electric potential and shear flow enhance the effective diffusion coefficients of each species in channel domains.  The exact solutions of the effective equation with certain special parameters, as well as the asymptotic analyses for ions with large diffusivity discrepancies, are presented. Furthermore, there are several interesting properties of the effective equation. First, it is a generalization of the Taylor dispersion, with a nonlinear diffusion tensor replacing the scalar diffusion coefficient. Second, the effective equation exhibits a scaling relation, revealing that the system with a weak flow is equivalent to the system with a strong flow under scaled  physical parameters. Third, in the case of injecting an electrolyte solution into a channel containing well-mixed buffer solutions or electrolyte solutions with the same ion species, if the concentration of the injected solution is lower than that of the pre-existing solution, the effective equation simplifies to a multidimensional diffusion equation. However, when introducing the electrolyte solution into a channel filled with deionized water, the ion-electric interaction results in several phenomena not present in the advection-diffusion equation,  including upstream migration of some species, spontaneous separation of ions, and non-monotonic dependence of the effective diffusivity on P\'eclet numbers. Last, the dependence of effective diffusivity on concentration and ion diffusivity suggests a method to infer the concentration ratio of each component and ion diffusivity by measuring the effective diffusivity.

\end{abstract}


\section{Introduction }
Fluid flow plays an important role in the transport of solutes. When a solute is transported in a fluid through a narrow tube or channel, the interaction of fluid flow and molecular diffusion causes the solute to spread out and become more dispersed as it travels down the tube.  This effect is known as Taylor dispersion, named after G. I. Taylor, who first investigated the phenomenon in  (\cite{taylor1953dispersion}). Since Taylor's seminal work, theoretical studies on Taylor dispersion  has exploded in many directions (\cite{aris1956dispersion,aris1960dispersion,chatwin1970approach,vedel2012transient,ding2022ergodicity}), and established applications in many disciplines such as  molecular diffusivity measurement (\cite{bello1994use,taladriz2019precision,leaist2017quinary}), chemical delivery in micro-channel (\cite{aminian2016boundaries,dutta2001dispersion}),  contaminant dispersion (\cite{chatwin1975longitudinal,smith1982contaminant,ngo2015higher}).

In an electrolyte solution, the electric current is carried by the dissolved ions. The electric field exerts significant body forces on the ions, affecting their fluxes, which is another key factor in mass transfer. Even in the absence of an external electric field, where the electroneutrality and zero current conditions are met, it is necessary to consider ion-electric interaction in multispecies electrolyte solutions because dissolved ions have different diffusivities. To maintain electroneutrality, the faster-moving ion is slowed down, creating a balance between positive and negative charges.
For example, sodium fluorescein is a commonly used tracer in fluid experiments, and its self-diffusion coefficient in water has been measured experimentally by several authors to be around $4- 5\times 10^{-6}$  cm$^{2}$s$^{-1}$(\cite{casalini2011diffusion}). However,  in a sodium chloride stratified fluid, the diffusion coefficient of sodium fluorescein could exhibit a significant increase, reaching values of   $8- 9 \times 10^{-6}$ cm$^{2}$s$^{-1}$ (\cite{ding2021enhanced}).

The system involves fluid flow, electric field, and diffusion can be well-described by   the advection-Nernst-Planck equation (\cite{deen1998analysis,lyklema2005fundamentals,cussler2013multicomponent}) . Many recent studies show that the transport of multiple electrolytes exhibits different properties compared with the transport of a single binary electrolyte (\cite{gupta2019diffusion,hosokawa2011development,liu2011multispecies}). When dealing with two different ion species, the nonlinear governing equation can be reduced to the advection-diffusion equation (\cite{deen1998analysis}), allowing for simplified analysis. However, when dealing with more than two different ion species, the complexity of the nonlinear governing equation prohibits simplification to the advection-diffusion equation, necessitating a comprehensive consideration of the electro-diffusive process to accurately describe the system's behavior.

Understanding how fluid flow, electric potential, and diffusion interact in multispecies electrolyte solutions is essential for accurately measuring mutual diffusion  (\cite{leaist1993diffusion,price1988theory,ribeiro2019coupled,rodrigo2022ternary,rodrigo2021coupled}), as well as for simulating the system with stratified fluids (\cite{poisson1983diffusion,ben1972diffusion,yuan1974diffusion,ding2021enhanced,ding2023dispersion}), and for controlling diffusiophoresis (\cite{ault2017diffusiophoresis,alessio2022diffusioosmosis}) and modeling isotachophoresis  (\cite{bhattacharyya2013sample,gopmandal2015effects,ganor2015diffusion}) and  chromatography (\cite{biagioni2022shape}). Despite its importance, the interplay between these three factors has not been extensively studied in the literature, creating a knowledge gap. The main goal of this study is to fill this gap by presenting a comprehensive investigation of this interplay.

To this end, we use homogenization methods to derive an effective equation that is valid at the diffusion time scale for the advection-Nernst-Planck equation in a channel with arbitrary cross-sectional geometry. In addition, the resulting effective equation depends only on the longitudinal variable of the channel, and provides a more tractable approximation for analyzing mass transfer which captures the combined effects of flow advection and ion-electric interaction. Our analysis of the effective equation shows that the variance of the concentration distribution asymptotically increases linearly with time, and we demonstrate that the effective diffusivity can be efficiently calculated via the self-similar solution of the effective equation.  Effective diffusivity is a critical parameter for understanding the mass transfer and guiding the designing of microfluidic devices (\cite{dutta2001dispersion}), and we show that it can also be used to infer the concentration ratio of each component and ion diffusivity in multispecies electrolyte solutions. We demonstrate that the effective equation exhibits a reciprocal property, namely, the system without flow is mathematically equivalent to the system with a strong flow and scaled physical parameters.  We derive the self-similarity solution of the effective equation and present asymptotic analyses for ions with large diffusivity discrepancies. Moreover, we find that the nonlinear effective equation can be approximated by a diffusion equation with mutual diffusion coefficients when the background concentration is nonzero, consistent with previous studies (\cite{rodrigo2022ternary}). 

To complement our analytical results, we conduct numerical simulations to explore the behavior of multispecies electrolyte solutions under different flow and electric field conditions, validating our analytical results. Our simulations reveal several interesting properties arising from the nonlinearity of the advection-Nernst-Planck equation, such as upstream migration of some species, separation of ions depending on the flow strength, the presence of highly non-Gaussian and bimodal shape of concentration distribution and a non-monotonic dependence of the effective diffusivity on P\'eclet numbers.

The paper is organized as follows: In Section \ref{sec:Governing equation and effective diffusivity}, we introduce the governing equations for the transport of multispecies electrolyte solutions in channel domains and provide a comprehensive overview of effective diffusivity. Section \ref{sec:Effective equation} presents the derivation of the effective equation for the advection-Nernst-Planck equation at long times using homogenization methods. In Subsection \ref{sec:Effective equation for some shear flows}, we outline the effective equation for specific shear flows in parallel-plate channel domains and circular pipes. Subsection \ref{eq:Self-similar solution of the effective equation} discusses the self-similarity solution for different types of initial conditions and presents the formula for calculating the effective diffusivity using this solution. Subsection \ref{sec:Comparison to the Taylor dispersion} compares our results with those of Taylor dispersion and highlights the reciprocal property exhibited by the effective equation. In Section \ref{sec:Analytical result}, we provide the exact solution of the effective equation for certain parameter combinations and analyze cases with significant differences in ion diffusivity. Section \ref{sec:Examples and numerical tests} validates our analytical results through numerical simulations and explores intriguing phenomena resulting from ion-electric interactions. Finally, in Section \ref{sec:Conclusion and discussion}, we summarize our findings and discuss potential avenues for future research.

 \section{Governing equation and effective diffusivity}
 \label{sec:Governing equation and effective diffusivity}
 \subsection{Advection-Nernst-Planck equation}
 \begin{figure}
  \centering
    \includegraphics[width=0.46\linewidth]{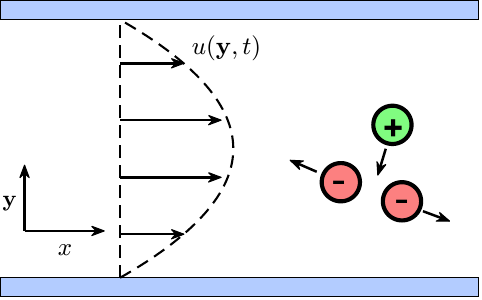}
  \hfill
  \caption[]
  {The schematic shows the setup for the special case of a quadratic shear flow in the parallel-plate channel domain. A multispecies electrolyte in water exists in the form of dissolved ions. Ions of like charges repel, while ions of opposite charges attract, due to electrostatic forces. The interplay between the flow and the ion-electric interactions plays a crucial role in determining the behavior and dynamics of the system.
  }
  \label{fig:ShearflowSchematic}
\end{figure}

We consider the electrolyte solution transport in a channel domain: $(x, \mathbf{y}) \in \mathbb{R}\times \Omega$, where the $x$-direction is the longitudinal direction of the channel and $\Omega \subset \mathbb{R}^{d}$ stands for the cross-section of the channel.  $\mathbf{n}$ is the outward normal vector of the boundary, $\mathbb{R} \times \partial\Omega$, where $\partial\Omega$ is the boundary of $\Omega$.  Some practical examples of the channel boundary geometry includes the parallel-plate channel $\Omega= \left\{ y| y \in [0,L_{y}]  \right\}$ (sketched in figure \ref{fig:ShearflowSchematic}), the circular pipe $\Omega=\left\{ \mathbf{y}| \mathbf{y}^{2}\leq L_{y}^{2} \right\}$, the rectangular duct $\Omega=\left\{ \mathbf{y}| \mathbf{y} \in [0,L_{y}]\times[0,H_{y}] \right\}$, and bowed rectangular channels (\cite{lee2021dispersion}).

Denote the concentration and valence of $i$-th species of ion as $c_{i} (x,\mathbf{y},t)$ and $z_i$, respectively. The concentration evolution of $n$ ion species under the shear flow advection and ionic interaction can be modeled by the Nernst-Planck equation (see section 11.7 in \cite{deen1998analysis}, or \cite{Maex2013Nernst}),
\begin{equation}\label{eq:Advection-Nernst-Planck equation}
\begin{aligned}
  &\partial_{t}c_{i}+\nabla \cdot \left( \mathbf{u} c_{i} \right)=\kappa_{i} \Delta c_{i}+ \frac{\kappa_{i} z_{i} e }{k_{B} T} \nabla \cdot(c_{i} \nabla \phi),\; c_{i} (x,\mathbf{y},0)=c_{I,i} \left( \frac{x}{L_{x}} \right), \; i=1,\hdots,n,
\end{aligned}
\end{equation}
where $\kappa_{i}$ is the diffusivity of the $i$-th species of ion, $\phi (x,\mathbf{y},t) $ is the electric potential, $e$ is the elementary charge, $k_B$ is the Boltzmann constant and $T$ is the temperature.  $c_{I,i}$ is the initial condition of the $i$-th species of ion. $L_{x}$ is the characteristic length scale  of the initial condition. The second term on the left-hand side of equation \eqref{eq:Advection-Nernst-Planck equation} describes the fluid flow advection.  The first term on the right-hand side of equation \eqref{eq:Advection-Nernst-Planck equation} describes the ion diffusive motion, while the second term represents the electromigration in response to the local electric field.

We assume the electrolyte solutions are passively advected by a prescribed velocity field which takes the form $\mathbf{u} = (u (\mathbf{y},t), 0, \dots, 0)$. The function $u (\mathbf{y},t)$ vanishes on the boundary wall and exhibits periodic time-varying behavior with a period of $L_{t}$. While steady pressure-driven flow is common in many applications (\cite{leaist1993diffusion,price1988theory,rodrigo2021coupled}), we maintain the general form and time-dependence of the flow to ensure the theoretical framework's applicability to various scenarios, including blood flow (\cite{marbach2019active}), and scalar intermittency (\cite{majda1999simplified,camassa2021persisting}).
We impose the no-flux boundary condition for the concentration fields of the ion species $\left. \mathbf{n}\cdot (\kappa_{i}\nabla c_{i} +  \frac{\kappa_{i} z_{i} e }{k_{B} T} c_{i} \nabla \phi) \right|_{\mathbb{R}\times \partial\Omega }=0$.

Now there are $n$ conservation equations for $n$ concentration fields and an unknown electric potential $\phi$. An additional Poisson equation can be derived from Gauss' law, which is one of Maxwell's equations of electricity and magnetism. When combined with the Nernst-Planck equation, it forms the Poisson-Nernst-Planck system (\cite{schmuck2015homogenization}).  However, this work focuses on the case in the absence of an external electric field. The net charge density is zero almost everywhere. We consider two alternative equations that serve as reasonable approximations of the Poisson equation in this setting. The first additional equation arises from the electroneutrality condition   $\sum\limits_{i=1}^{n} z_{i}c_{i}=0$. The second condition is the zero electric current condition, given by $\mathbf{0}= \sum\limits_{i=1}^{n}z_{i}\mathbf{J}_{i}=\sum\limits_{i=1}^{n}z_{i} \left( \mathbf{u} c_{i}-  \kappa_{i} \nabla c_{i}- \frac{\kappa_{i}z_{i} e }{k_B T} c_{i} \nabla \phi \right) $, which is commonly used in the literature when there is no external electric field  (\cite{ben1972diffusion,tournassat2020solving,gupta2019diffusion,tabrizinejadas2021validity}). Moreover, for the electroneutrality initial data, the zero electric current condition ensures that the  electroneutrality condition is always true (see \cite{boudreau2004multicomponent}).

Using the zero electric current condition,  the gradient of the electric potential can be expressed in terms of ion concentrations
\begin{equation}\label{eq:diffusion induced electric potential}
\begin{aligned}
\frac{e }{k_BT} \nabla \phi= \frac{\sum\limits_{i=1}^{n}z_{i} \left(\mathbf{u}c_{i}
- \kappa_{i} \nabla c_{i} \right)}{\sum\limits_{i=1}^{n}z_{i}^{2}  \kappa_{i} c_{i} }=\frac{\sum\limits_{i=1}^{n-1}(\kappa_{n}- \kappa_{i})  z_{i} \nabla c_{i} }{\sum\limits_{i=1}^{n-1} (z_{i} \kappa_{i}-z_{n}\kappa_{n})z_{i}c_{i} },
\end{aligned}
\end{equation}
where the second step follows the electroneutrality condition. Equation \eqref{eq:diffusion induced electric potential} shows that the electric potential gradient is induced  by the difference in ion diffusivities. When all diffusivities take the same value, the gradient of the diffusion-induced potential becomes zero, and equation \eqref{eq:Advection-Nernst-Planck equation} reduces to the advection-diffusion equation. When there is a  difference in diffusivities, substituting equation \eqref{eq:diffusion induced electric potential} into Nernst-Planck equation \eqref{eq:Advection-Nernst-Planck equation} yields the equation that will be mainly used in this study
\begin{equation}\label{eq:Advection-Nernst-Planck equation no potential}
\begin{aligned}
  &\partial_{t}c_{i}+u (\mathbf{y},t)\partial_{x} c_{i}=\kappa_{i} \Delta c_{i}+ \kappa_{i} z_{i}  \nabla \cdot \left( \frac{ c_{i} \sum\limits_{j=1}^{n-1} (\kappa_{n}- \kappa_{j}) z_{j} \nabla c_{j} }{\sum\limits_{j=1}^{n-1}(z_{j} \kappa_{j}-z_{n}\kappa_{n}) z_{j}c_{j}  }  \right), \quad i=1,\hdots,n-1.\\
 \end{aligned}
\end{equation}
The system of equations  exhibits an interesting  scaling property, wherein any solution multiplied by a constant remains a valid solution to the system. Furthermore, if all valences are multiplied by a constant, the original solution of the system remains a solution to the system with the new valences.

\begin{figure}
  \centering
    \includegraphics[width=0.6\linewidth]{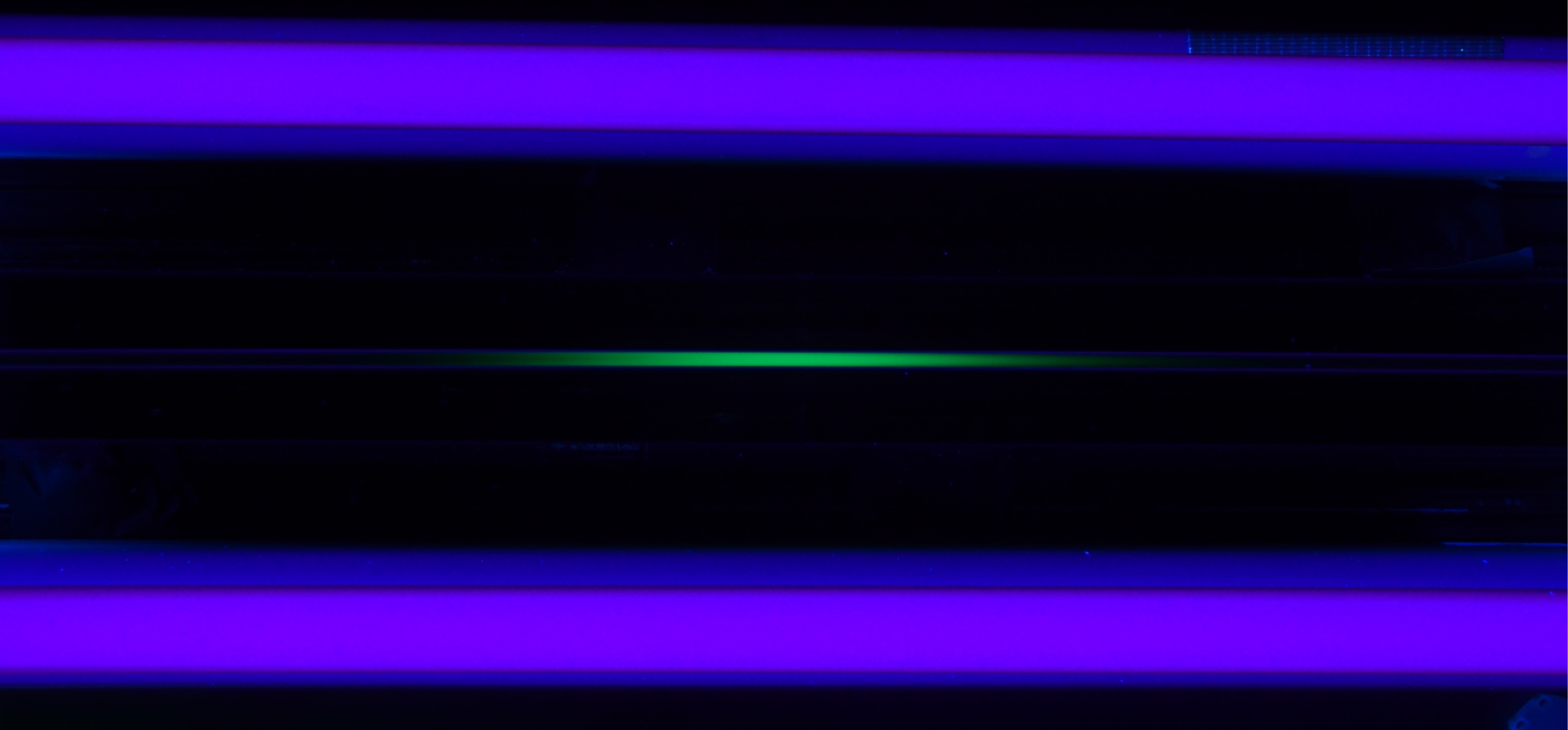}
  \hfill
  \caption[]
  {  This experimental photo shows the advection of a sodium fluorescein solution in a pipe through pressure-driven flow. The purple regions in the photo represent the presence of ultraviolet tube lights. Upon exposure to these lights, the sodium fluorescein solution emits a vibrant green light. The experimental setup includes a pump attached to the left end of the pipe, which generates a steady pressure-driven flow to transport the sodium fluorescein solution towards the right.   For the detailed procedure and setup of the laminar channel flow experiments, we refer to the article \cite{aminian2018diffusion}.}
  \label{fig:sodium fluorescein pipe flow}
\end{figure}

We proceed by considering a combination of typical experimental physical parameters, aiming to identify the dominant factors in the problem and facilitate perturbation analysis. The diffusivity of the solute is around $10^{-5}$ cm$^{2}$/s. The length scale of concentration, denoted as $L_{x}$, ranges from millimeters to centimeters. Meanwhile, the channel width, denoted as $L_{y}$, spans from micrometers to millimeters. As depicted in Figure \ref{fig:sodium fluorescein pipe flow}, the advection caused by flow and the diffusion contribute to the phenomenon, leading to the condition $L_{x}\gg L_{y}$. The characteristic fluid velocity varies from millimeters per second to centimeters per second. Additionally, apart from microfluidic experiments, our study also have implications for blood flow scenarios, where a rich variety of electrolytes are present. Depending on the type of blood vessel, their radii typically range from 10 to 200 micrometers, and blood velocities vary from 0.1 cm/s to 20 cm/s.

When an object's surface is exposed to a fluid, two parallel layers of charge surrounding the object appear. Specifically, under a strong applied electric field, electro-osmotic flow occurs (\cite{ghosal2012electromigration}), which has numerous applications in microfluidics. One might question whether the assumptions of electroneutrality and zero current still hold in the presence of surface changes. However, in this study, we can neglect the effects of surface charge due to the following reasons. Firstly, the characteristic thickness of the double layer, known as the Debye length, is typically on the order of nanometers (\cite{hashemi2018oscillating}). This is significantly smaller than the characteristic width of the micro-channel, which ranges from micrometers to millimeters. Secondly, in the absence of an external electric field, electro-osmotic flow is negligible compared to the fluid flow imposed by other factors, such as the pressure-driven flow produced by the pump.

\subsection{Effective diffusivity}
As demonstrated by many studies (\cite{taylor1953dispersion,aris1956dispersion,chatwin1970approach,ding2022determinism}), the solution of the advection-diffusion equation in the channel domain converges to a Gaussian distribution function at long times. To model this behavior, one can use a diffusion equation with an enhanced effective diffusion coefficient.  Therefore, understanding the dependence of the effective diffusion coefficient on the flow conditions, channel geometries, and ion physical parameters is important for optimizing microfluidic device performance, either enhancing or reducing mixing (\cite{dutta2001dispersion,aminian2016boundaries,aminian2015squaring}).

The precise definition of the effective diffusion coefficient depends on the initial condition. In this study, we consider three types of initial conditions. The first type of initial condition is an integrable function that vanishes at infinity, such as $c_{I,i} (x) = \frac {2}{\sqrt {\pi }}e^{-x^{2}}$. This type of initial condition can be used to model the delivery of chemicals with a finite volume in a capillary tube (see \cite{aminian2016boundaries}). In the second type of initial condition, the concentration field can be expressed as $c_{i} (x,\mathbf{y},t) =c_{i} (\infty)+ \tilde{c}_{i} (x,\mathbf{y},t)$, where $c_{i} (\infty)$ is a constant representing the background concentration, and $\tilde{c}_{i}$ is an integrable function representing the deviation from the background concentration, for example, $c_{I,i} (x) =1+\frac {2}{\sqrt {\pi }}e^{-x^{2}}$. In many experimental studies (e.g., \cite{leaist2001coupled,leaist2017quinary}), the pipe is filled with buffer solutions. In such cases, $c_{i}(\infty)>0$, and $\tilde{c}_i$ can take negative values as long as $c_i$ remains non-negative. In the third type of initial condition, the concentration field tends to a constant value at infinity, but the values at positive and negative infinity can be different, such as $c_{I,i} (x) = \mathrm {erf} (x) =\frac {2}{\sqrt {\pi }}\int _{0}^{x}e^{-t^{2}}\mathrm {d} t$, which can be used to model the continuous injection of a solution with a constant concentration into the channel domain (see \cite{taylor1953dispersion}). The solutions of the equations with these three types of initial conditions exhibit different long-time asymptotic properties, and therefore we have treated them separately in our analysis.

For the first type of initial condition, the effective longitudinal effective diffusivity is given by 
\begin{equation}\label{eq:effectiveDiffusivityDefinition}
\kappa_{\mathrm{eff},i}=   \lim\limits_{t\rightarrow \infty}\frac{\mathrm{Var} (\bar{c}_{i})}{2t \int\limits_{-\infty}^{\infty} \bar{c}_{i}\mathrm{d} x }= \lim\limits_{t\rightarrow \infty}\frac{\partial_{t}\mathrm{Var} (\bar{c}_{i})}{2 \int\limits_{-\infty}^{\infty} \bar{c}_{i}\mathrm{d} x } ,
\end{equation}
where  $ \bar{c}_{i} (x,t) = \frac{1}{\left| \Omega\right|}  \int\limits_{\Omega}^{}c_{i} (x,\mathbf{y},t) \mathrm{d}\mathbf{y}$ is the cross-sectional average of the scalar field $c_{i}$.  $\left| \Omega \right|$  is the area of $\Omega$. $\mathrm{Var} (\bar{c}_{i})= \int\limits_{-\infty}^{\infty} \bar{c}_{i} x^{2}\mathrm{d}x - \left( \int\limits_{-\infty}^{\infty} \bar{c}_{i} x\mathrm{d} x \right)^{2}$ is the variance of the cross-sectional averaged concentration field $ \bar{c}_{i}$. In other words, the asymptotics of the variance is given by
\begin{equation}\label{eq:variance asymptotics}
\begin{aligned}
\mathrm{Var} (\bar{c}_{i}) \approx \mathrm{Var} (\bar{c}_{I,i})+ 2 t \kappa_{\mathrm{eff}} \int\limits_{-\infty}^{\infty} \bar{c}_{i}\mathrm{d} x, \quad t \rightarrow \infty.
\end{aligned}
\end{equation}

For the second type of initial condition, where the background ion concentration is nonzero, $c_i$ is not integrable. One can define  the effective longitudinal effective diffusivity via the perturbed concentration
\begin{equation}\label{eq:effectiveDiffusivityDefinition perturbed}
\kappa_{\mathrm{eff},i}=   \lim\limits_{t\rightarrow \infty}\frac{\mathrm{Var} (\bar{c}_{i}-c_{i} (\infty))}{2t \int\limits_{-\infty}^{\infty} \bar{c}_{i}-c_{i} (\infty)\mathrm{d} x }= \lim\limits_{t\rightarrow \infty}\frac{\partial_{t}\mathrm{Var} (\bar{c}_{i}-c_{i} (\infty))}{2 \int\limits_{-\infty}^{\infty} \bar{c}_{i}-c_{i} (\infty)\mathrm{d} x }.
\end{equation}

The solution with  the third of initial condition is also not integrable, but we can investigate its derivative,
\begin{equation}
\kappa_{\mathrm{eff},i}=   \lim\limits_{t\rightarrow \infty}\frac{\mathrm{Var} (\partial_{x}\bar{c}_{i})}{2t (\bar{c}_{i} (\infty) -c_{i} (-\infty)) }=  \lim\limits_{t\rightarrow \infty}\frac{\partial_{t} \mathrm{Var} (\partial_{x}\bar{c}_{i})}{2 (\bar{c}_{i} (\infty) -c_{i} (-\infty))}.
\end{equation}

Although the diffusion-induced electric potential may cause the concentration field to deviate from a Gaussian distribution function, we are still interested in computing the effective diffusivity for several reasons. First, when the electric potential is weak and the background concentration is nonzero, the solution can be reasonably approximated by a Gaussian distribution function or error function. Second, as the time approaches infinity, the longitudinal variance of the concentration field  increases linearly, ensuring that the effective diffusivity remains well-defined quantity for characterizing the system. Third, by examining the relationship between effective diffusivity and other physical paramters, e.g., ion diffusivity, one can devise an experimental method for measuring the latter.

\section{Effective equation}
\label{sec:Effective equation}
It is possible to develop a simplified model that depends only on the longitudinal variable and time, given that the length scale in the longitudinal direction of the channel domain is significantly larger than the length scale in the transverse direction (as shown in figure \ref{fig:sodium fluorescein pipe flow}). By simplifying the model in this way, one can reduce the computational complexity of the problem while retaining the relevant physical phenomena without compromising the key features of interest. The homogenization method is a widely used method to acheive this goal, especially for the linear advection-diffusion problem (\cite{camassa2010exact,wu2014approach} ). Here, we will employ the homogenization method to derive the effective equation for the nonlinear equation \eqref{eq:Advection-Nernst-Planck equation}.

\subsection{Homogenization  method}\label{sec:Homogenization  method}
The first step is to non-dimensionalize the equation, which helps identify the dominant terms. The change of variables for the nondimensionalization is
\begin{equation}
\begin{aligned}
&L_{x}x'=x,\quad L_{y}\mathbf{y}'=\mathbf{y}, \quad \epsilon= \frac{L_{y}}{L_{x}}, \quad \frac{L_{y}^{2}}{\tilde{\kappa} \epsilon^2}t'=t,\\
&\tilde{c}c_{i}'=c_{i},  \quad \frac{e }{k_BT}   \phi'=\phi,\quad  Uu' \left( \mathbf{y}',\frac{t'}{\epsilon^{2}} \right)= u(\mathbf{y},t),
\end{aligned}
\end{equation}
where $\tilde{c}$ is the characteristic concentration, $\tilde{\kappa}$ is the characteristic diffusivity. One can drop the primes without confusion and obtain the non-dimensionalized equation,
\begin{equation}\label{eq:Advection-Nernst-Planck equation non dimensional}
\begin{aligned}
  &\partial_{t}c_{i}+\frac{\mathrm{Pe}u  \left( \mathbf{y}, \frac{t}{\epsilon^2} \right)}{\epsilon} \partial_{x} c_{i}=\kappa_{i} \partial_{x}\left( \partial_{x}c_{i}+z_{i}c_{i}\partial_{x}\phi \right)+ \frac{\kappa_{i} }{\epsilon^{2}}\nabla_{\mathbf{y}} \cdot \left(\nabla_{\mathbf{y}}c_{i}+z_{i} c_{i}\nabla_{\mathbf{y}}\phi \right),\\
  & \left. c_{i} \right|_{t=0 }=c_{I,i} \left( x \right),\; \left. \mathbf{n}\cdot  \kappa_{i} (\partial_{x}c_{i}+ \frac{1}{\epsilon}\nabla_{\mathbf{y}} c_{i} +  z_{i} c_{i}(\partial_{x}\phi + \frac{1}{\epsilon}\nabla_{\mathbf{y}} \phi)) \right|_{\mathbb{R}\times \partial\Omega }=0, \;i=1,\hdots,n-1,\\
  &\nabla \phi= \frac{\sum\limits_{i=1}^{n-1} (\kappa_{n}- \kappa_{i}) z_{i} \nabla c_{i} }{\sum\limits_{i=1}^{n-1} (z_{i} \kappa_{i}-z_{n}\kappa_{n})z_{i}c_{i} },
\end{aligned}
\end{equation}
where $\nabla_{\mathbf{y}}= \left( \partial_{y_{1}}, \hdots, \partial_{y_{d}} \right)$,  $ \mathrm{Pe}=\frac{L_{y}U}{\tilde{\kappa}}$ is the P\'{e}clet number and  $u (\mathbf{y},t)$ has a temporal period $\tilde{L}_{t}=\kappa L_{t}/L_{y}^{2}$. It is convenient to introduce two different scales in time: $t$ (slow), $\tau= {t}/{\epsilon^2}$ (fast). Consequently, the differential operators in time will be replaced $\partial_{t}\rightarrow    \partial_{t}+\frac{1}{\epsilon^{2}}\partial_{\tau}$  and the equation becomes
\begin{equation}\label{eq:non-dimensional advection-Nernst-Planck equation multiscale}
\begin{aligned}
  &\partial_{t}c_{i}+\frac{\partial_{\tau}c_{i}}{\epsilon^{2}}+\frac{\mathrm{Pe}}{\epsilon} u  \left( \mathbf{y}, \tau \right)\partial_{x} c_{i}=\kappa_{i} \partial_{x}\left( \partial_{x}c_{i}+z_{i}c_{i}\partial_{x}\phi \right)+ \frac{\kappa_{i} }{\epsilon^{2}}\nabla_{\mathbf{y}} \cdot \left(\nabla_{\mathbf{y}}c_{i}+z_{i} c_{i}\nabla_{\mathbf{y}}\phi \right).\\
\end{aligned}
\end{equation}

Notice that the equation is invariant under the translation in $x$. For convenience, one can consider applying the Galilean transformation $\tilde{x}=x- \left\langle u (\mathbf{y},\tau) \right\rangle_{\mathbf{y},\tau} t$ so that the resulting new shear flow $\tilde{u}=u- \left\langle u (\mathbf{y},\tau) \right\rangle_{\mathbf{y},\tau}   $ has a zero average, where  the average of a function is defined as $\left\langle f (\mathbf{y},\tau) \right\rangle_{\mathbf{y},t}=\frac{1}{|\Omega|\tilde{L}_{t}}\int\limits_{\Omega}^{} \int\limits_{0}^{\tilde{L}_{t}} f (\mathbf{y},\tau) \mathrm{d} \mathbf{y} \mathrm{d} \tau$.

Assuming the asymptotic expansion of $c_{i}$ in the limit $\epsilon\rightarrow 0$ is
\begin{equation}
\begin{aligned}
  &c_{i}(x,\mathbf{y},t)=c_{i,0}(x,y,t,\tau)+\epsilon c_{i,1}(x,y,t,\tau)+\epsilon^{2}c_{i,2}(x,y,t,\tau)+\mathcal{O}(\epsilon^3).\\
\end{aligned}
\end{equation}
Substituting the asymptotic expansion of $c_i$ into the formula for $\phi$ and using the Taylor expansion yield the asymptotic expansion of $\phi$, i.e., $\phi=\phi_{0}+\epsilon \phi_{1}+\epsilon^{2} \phi_{2} +\mathcal{O} (\epsilon^{3})$. In particular, the gradient of the first two coefficients are given by  
\begin{equation}\label{eq:potential expansion coefficeint}
\begin{aligned}
  & \nabla \phi_{0}=  \frac{\sum\limits_{i=1}^{n-1} (\kappa_{n}- \kappa_{i}) z_{i} \nabla c_{i,0} }{\sum\limits_{i=1}^{n-1} (z_{i} \kappa_{i}-z_{n}\kappa_{n})z_{i}c_{i,0} },\\
    & \nabla \phi_{1}= \frac{\sum\limits_{i=1}^{n-1} (\kappa_{n}- \kappa_{i})z_{i} \nabla c_{i,1} }{\sum\limits_{i=1}^{n-1} (z_{i} \kappa_{i}-z_{n}\kappa_{n}) z_{i}c_{i,0} }
  +  \frac{\left(\sum\limits_{i=1}^{n-1}(\kappa_{n}- \kappa_{i})  z_{i} \nabla c_{i,0}  \right)   \left( \sum\limits_{i=1}^{n-1} (z_{i} \kappa_{i}-z_{n}\kappa_{n})z_{i}c_{i,1} \right)}{ \left(\sum\limits_{i=1}^{n-1}(z_{i} \kappa_{i}-z_{n}\kappa_{n})z_{i}c_{i,0}    \right)^{2}}.
\end{aligned}
\end{equation}
Substituting the expansion of $c_i$ and $\nabla \phi$ into equation \eqref{eq:non-dimensional advection-Nernst-Planck equation multiscale} leads to an equation involving the power series of $\epsilon$. Since the equation holds for arbitrarily small $\epsilon$,  the coefficient of each power of $\epsilon$ should be zero, which yields a hierarchy of equations of $c_{i,k}$.

 Grouping all term of order $\mathcal{O}(\epsilon^{-2})$ and setting the coefficient to be zero  yield the equation 
\begin{equation}
\partial_{\tau}c_{i,0}=\kappa_{i}\nabla_{\mathbf{y}}\cdot \left( \nabla_{\mathbf{y}}c_{i,0}+z_{i} c_{i,0}\nabla_{\mathbf{y}}\phi_{0} \right), \quad  \left. c_{i,0} \right|_{t=0,\tau=0 }=c_{I,i} \left( x \right).
\end{equation}
The initial condition is a function of the variable $x$ only, which means  that $c_{i,0}(x,\mathbf{y},t,\tau)=c_{i,0}(x,t)$, $i=1, \hdots,n$.  Consequently, the evolution equation for $c_{i,0}$ provides the desired approximation. The goal of this homogenization calculation is to derive this equation.

 Grouping all term of order $\mathcal{O}(\epsilon^{-1})$  yields the equation 
 \begin{equation}\label{eq:multiscale expansion epsilon -1}
\begin{aligned}
 &\partial_{\tau}c_{i,1}+\mathrm{Pe} u(\mathbf{y},\tau) \partial_{x} c_{i,0}=\kappa_{i}\nabla_{\mathbf{y}}\cdot \left(\nabla_{\mathbf{y}}c_{i,1}+z_{i} c_{i,1}\nabla_{\mathbf{y}}\phi_{0}+z_{i} c_{i,0}\nabla_{\mathbf{y}}\phi_{1}  \right),\\
\end{aligned}
\end{equation}
with the initial condition $\left. c_{i,1} \right|_{t=0,\tau=0 }=0$ and the no-flux boundary condition $\left. \mathbf{n}\cdot \left( \nabla_{\mathbf{y}}c_{i,1}+z_{i} c_{i,1}\nabla_{\mathbf{y}}\phi_{0}+z_{i} c_{i,0}\nabla_{\mathbf{y}}\phi_{1}   \right)\right|_{\mathbb{R}\times \partial\Omega }=0$, $i=1,\hdots,n-1$.
Since $c_{i,0}$ is independent of $\mathbf{y}$, equation \eqref{eq:potential expansion coefficeint} implies
\begin{equation}
\begin{aligned}
 \nabla_{\mathbf{y}}\phi_{0}=0, \quad  \nabla_{\mathbf{y}} \phi_{1}=\frac{\sum\limits_{i=1}^{n-1} (\kappa_{n}- \kappa_{i})z_{i} \nabla_{\mathbf{y}} c_{i,1} }{\sum\limits_{i=1}^{n-1} (z_{i} \kappa_{i}-z_{n}\kappa_{n}) z_{i}c_{i,0} } .
  \end{aligned}
\end{equation}
Therefore,  equation \eqref{eq:multiscale expansion epsilon -1} is a linear equation of $c_{i,1}$,
\begin{equation}\label{eq:multiscale expansion epsilon -1 simplified}
\begin{aligned}
  &\partial_{\tau}\mathbf{c}_{1}+\mathrm{Pe} u(\mathbf{y},\tau) \partial_{x} \mathbf{c}_{0}= \mathbf{D} (\mathbf{c}_{0}) \Delta_{\mathbf{y}}\mathbf{c}_{1},\\
  &\mathbf{D}=
  \begin{bmatrix}
    \kappa_{1}&\hdots&0\\
    &\hdots&\\
    0&\hdots&\kappa_{n-1}
  \end{bmatrix}
-\begin{bmatrix}
  \kappa_{1}z_{1}c_{1,0}\\
  \hdots\\
  \kappa_{n-1}z_{n-1}c_{n-1,0}
\end{bmatrix} \frac{
  \begin{bmatrix}
(\kappa_{1}-\kappa_{n})z_{1},\hdots, (\kappa_{n-1}-\kappa_{n})z_{n-1}
  \end{bmatrix}
}{\sum\limits_{i=1}^{n-1} (z_{i} \kappa_{i}-z_{n}\kappa_{n}) z_{i}c_{i,0}} ,\\
\end{aligned}
\end{equation}
where
$\mathbf{c}_{0}= (c_{1,0},\hdots, c_{n-1,0})$, $\mathbf{c}_{1}= (c_{1,1},\hdots, c_{n-1,1})$,  $\Delta_{\mathbf{y}}= \sum\limits_{i=1}^{d}\partial_{y_{i}}^{2}$, and $\Delta_{\mathbf{y}}\mathbf{c}_{1}= (\Delta_{\mathbf{y}}c_{1,1},\hdots, \Delta_{\mathbf{y}}c_{n-1,1})$. Hence, the matrix \textbf{D} is  the difference of a diagonal matrix and an outer product of two vectors.

For unsteady shear flow $u (\mathbf{y},t)$, the solution of this diffusion equation can be expressed as
\begin{equation}\label{eq:c1 expression unsteady}
\begin{aligned}
& \mathbf{c}_{1}= \mathrm{Pe}(-\partial_{\tau}+\mathbf{D}\Delta_{\mathbf{y}})^{-1}\left( u\partial_{x}\mathbf{c}_{0} \right).   \\
\end{aligned}
\end{equation}
If the periodic unsteady shear flow admits a Fourier integral representation $u(\mathbf{y},t)= \left( 2\pi \right)^{-\frac{1}{2}}\int\limits_{\mathbb{R}}^{}  e^{\mathrm{i} tk}\hat{u}(\mathbf{y}, k) \mathrm{d}k$, then we can obtain the integral representation of $\mathbf{c}_{1}$:
\begin{equation}
\begin{aligned}
&\mathbf{c}_{1}=\mathrm{Pe} \left( 2\pi \right)^{-\frac{1}{2}}\int\limits_{\mathbb{R}}^{}  e^{\mathrm{i} tk} \hat{u}(\mathbf{y}, k)   (-\mathrm{i} k + \mathbf{D} \Delta_{\mathbf{y}})^{-1}\partial_{x}\mathbf{c}_{0} \mathrm{d} k. \\
\end{aligned}
\end{equation}

For the steady shear flow $u (\mathbf{y})$, the expression of $\mathbf{c}_{1}$ simplifies to 
\begin{equation}
\begin{aligned}
& \mathbf{c}_{1}= \mathrm{Pe}\Delta_{\mathbf{y}}^{-1}\left( u \right) \mathbf{D}^{-1}\partial_{x}\mathbf{c}_{0} ,   \\
\end{aligned}
\end{equation}
where the inverse of $\mathbf{D}$ is available from the Sherman--Morrison formula (see \cite{sherman1950adjustment})
\begin{equation}
  \begin{aligned}
      &\mathbf{D}^{-1}=
  \begin{bmatrix}
  \frac{1}{\kappa_{1}} &\hdots&0\\
    &\hdots&\\
    0&\hdots&\frac{1}{\kappa_{n-1}}
  \end{bmatrix} \left( 
    I_{n-1}+
    \begin{bmatrix}
  \kappa_{1}z_{1}c_{1,0}\\
  \hdots\\
  \kappa_{n-1}z_{n-1}c_{n-1,0}
\end{bmatrix} \frac{
  \begin{bmatrix}
\frac{(\kappa_{1}-\kappa_{n})z_{1}}{\kappa_{1}},\hdots, \frac{(\kappa_{n-1}-\kappa_{n})z_{n-1}}{\kappa_{n-1}}
  \end{bmatrix}
}{\kappa_{n}\sum\limits_{i=1}^{n-1}  (z_{i}-z_{n}) z_{i}c_{i,0}} \right) ,\\
\end{aligned}
\end{equation}
where $I_{n-1}$ is the $(n-1)\times (n-1)$ identity matrix.

Additionally concern is that whether  equation \eqref{eq:multiscale expansion epsilon -1 simplified} is solvable.  Fredholm solvability states that the linear equation $\mathcal{L}\Psi=f$ has a solution if and only if $\left\langle fg \right\rangle=0$ for any solution of equation $\mathcal{L}^{*}g=0$, where $\mathcal{L}^{*}$ is the adjoint operator of $\mathcal{L}$.  Here, the constant function solves the adjoint problem and the solvability condition of \eqref{eq:multiscale expansion epsilon -1 simplified} is guaranteed by the assumption that the average of flow is zero
\begin{equation}
\begin{aligned}
\left\langle  \mathrm{Pe} u(\mathbf{y},\tau) \partial_{x} \mathbf{c}_{0} \right\rangle_{\mathbf{y},\tau}=  \mathrm{Pe}\left\langle  u(\mathbf{y},\tau)\right\rangle_{\mathbf{y},\tau}\partial_{x} \mathbf{c}_{0} =0.
\end{aligned}
\end{equation}

 Grouping all $\mathcal{O}(\epsilon^{0})$ terms yields the equation
\begin{equation}
  \begin{aligned}
    &\partial_{t}c_{i,0}+\partial_{\tau}c_{i,2}+\mathrm{Pe} u  \left( \mathbf{y}, \tau \right)\partial_{x} c_{i,1}\\
    =&\kappa_{i} \partial_{x}\left( \partial_{x}c_{i,0}+z_{i}c_{i,0}\partial_{x}\phi_{0} \right)+ \kappa_{i} \nabla_{\mathbf{y}} \cdot \left(\nabla_{\mathbf{y}}c_{i,2}+z_{i} \left(  c_{i,2}\nabla_{\mathbf{y}}\phi_{0}+ 2c_{i,1}\nabla_{\mathbf{y}}\phi_{1}+ c_{i,0}\nabla_{\mathbf{y}}\phi_{2} \right) \right).\\
\end{aligned}
\end{equation}
In order to ensure the existence of a solution, the solvability condition requires the forcing term to have a zero average. When no-flux boundary conditions are imposed and the divergence theorem is applied, the average of the last term on the right-hand side of the above equation is shown to be zero. Therefore, the solvability condition can be expressed as
\begin{equation}
\begin{aligned}
&  \partial_{t}c_{i,0}+\mathrm{Pe}  \left\langle u  \left( \mathbf{y}, \tau \right)\partial_{x} c_{i,1} \right\rangle_{\mathbf{y},\tau}    =\kappa_{i} \partial_{x}\left( \partial_{x}c_{i,0}+z_{i}c_{i,0}\partial_{x}\phi_{0} \right).\\ 
\end{aligned}
\end{equation}
One can eliminate $c_{i,1}$ using equation \eqref{eq:c1 expression unsteady} and obtain the equation of $c_{i,0}$
\begin{equation}\label{eq:homogenization unsteady effective equation}
  \begin{aligned}
 &   \partial_{t}\mathbf{c}_{0}+\mathrm{Pe}^{2}  \partial_{x}\left\langle u   ( \mathbf{D}\Delta_{\mathbf{y}}-\partial_{\tau})^{-1}\left( u \partial_{x}\mathbf{c}_{0} \right)\right\rangle_{\mathbf{y},\tau}    =\partial_{x} (\mathbf{D} \partial_{x}\mathbf{c}_{0}),
  \end{aligned}
\end{equation}
where $\mathbf{D}$ is defined in equation \eqref{eq:multiscale expansion epsilon -1 simplified}.
For the steady shear flow, the equation reduces to
\begin{equation}\label{eq:homogenization steady effective equation}
  \begin{aligned}
    &    \partial_{t}\mathbf{c}_{0}+\mathrm{Pe}^{2}  \partial_{x}\left\langle u   \Delta_{\mathbf{y}}^{-1}u \right\rangle_{\mathbf{y},\tau} \mathbf{D}^{-1}\partial_{x}\mathbf{c}_{0}  =\partial_{x}(\mathbf{D} \partial_{x}\mathbf{c}_{0}).
  \end{aligned}
\end{equation}

The constant coefficient nonlinear equations \eqref{eq:homogenization steady effective equation} is an approximation of equation \eqref{eq:Advection-Nernst-Planck equation} in the limit of $\epsilon\rightarrow 0$, as well as at long times. It is worth noting that, as time elapses, the diffusion term in equation \eqref{eq:Advection-Nernst-Planck equation} smooths out the solution, which leads to an increase in the length scale of the solution $L_{x}$ and a decrease in the ratio $\epsilon=\frac{L_y}{L_x}$.

Finally, it should be noted that the homogenization calculation presented in this paper is not limited to the equation studied here. In fact, it can be applied to other nonlinear equations, including the one governing shear-enhanced diffusion in colloidal suspensions \cite{griffiths2012axial}. Moreover, this method offers a systematic way to obtain higher order approximations for these equations.

\subsection{Effective equation for some shear flows}\label{sec:Effective equation for some shear flows}
This section summarizes the effective equation derived by the homoginization method and presents explicit expression of the coefficient $\left\langle u   \Delta_{\mathbf{y}}^{-1}u \right\rangle_{\mathbf{y},\tau}$ in equation \eqref{eq:homogenization steady effective equation} for  some classical flows and the flow used in the numerical simulation.

The inversion of the Laplace operator in equation \eqref{eq:homogenization steady effective equation} depends on the domain geometry. In the parallel-plate channel domain, $\Omega= \left\{ y| y \in [0,1]  \right\}$, the formula is
\begin{equation}
\begin{aligned}
\Delta^{-1}u= \int\limits_0^y \int\limits_0^{s_{2}} u (s_{1})\mathrm{d}s_{1}\mathrm{d} s_{2}. 
\end{aligned}
\end{equation}

In the pipe geometry, $\Omega= \left\{ \mathbf{y}| |\mathbf{y}|\leq 1  \right\}$, the formula for an axisymmetric function $u (r)$,  $r=|\mathbf{y}|$  is 
\begin{equation}
\begin{aligned}
\Delta^{-1}u= \int\limits_0^r \frac{1}{s_{2}} \int\limits_0^{s_{2}} s_{1} u (s_{1})\mathrm{d}s_{1}\mathrm{d} s_{2}. 
\end{aligned}
\end{equation}

In the parallel-plate channel domain, the non-dimensionalized pressure-driven shear flow is $u=4(1 - y)y$, where the characteristic velocity is selected to be the maximum velocity. To use the conclusion in section \ref{sec:Homogenization  method}, one have to make a Galilean translation in the $x$-direction as mentioned earlier, so that the average shear over the transverse plane has mean zero. The shear flow in the new frame of reference is $u=4 \left( (1 - y)y + \frac{1}{6} \right)$. With this expression, $\left\langle u   \Delta_{\mathbf{y}}^{-1}u \right\rangle_{\mathbf{y},\tau}=\frac{-2 }{945}$ and equation \eqref{eq:homogenization steady effective equation} becomes
\begin{equation}\label{eq:homogenization steady effective equation parallel plate}
\begin{aligned}
    \partial_{t}\mathbf{c}_{0}  =\partial_{x} \left( \left( \mathbf{D}+\frac{\mathrm{2Pe}^{2}}{945} \mathbf{D}^{-1} \right) \partial_{x}\mathbf{c}_{0} \right).
\end{aligned}
\end{equation}

The numerical simulation presented in this paper  uses a simpler shear flow profile $u (y)=\cos (2 \pi y)$, which leads to $\left\langle u   \Delta_{\mathbf{y}}^{-1}u \right\rangle_{\mathbf{y},\tau}=\frac{-1}{8\pi^{2}}$. The corresponding effective equation is
\begin{equation}\label{eq:homogenization steady effective equation cos}
\begin{aligned}
    \partial_{t}\mathbf{c}_{0}  =\partial_{x} \left( \left( \mathbf{D}+\frac{\mathrm{Pe}^{2}}{8 \pi^{2}} \mathbf{D}^{-1} \right) \partial_{x}\mathbf{c}_{0} \right).
\end{aligned}
\end{equation}

In the pipe geometry, the non-dimensionalized pressure-driven shear flow in the mean velocity frame of reference is $u=\frac{1}{2}-r^{2}$, $r=|\mathbf{y}|$.  With this expression, $\left\langle u   \Delta_{\mathbf{y}}^{-1}u \right\rangle_{\mathbf{y},\tau}=\frac{-1}{192}$ and equation \eqref{eq:homogenization steady effective equation} becomes
\begin{equation}\label{eq:homogenization steady effective equation circular pipe}
\begin{aligned}
    \partial_{t}\mathbf{c}_{0}  =\partial_{x} \left( \left( \mathbf{D}+\frac{\mathrm{Pe}^{2}}{192} \mathbf{D}^{-1} \right) \partial_{x}\mathbf{c}_{0} \right).
\end{aligned}
\end{equation}

\subsection{Self-similar solution of the effective equation}
\label{eq:Self-similar solution of the effective equation}
Deriving the exact solution of the initial value problem \eqref{eq:homogenization unsteady effective equation} and \eqref{eq:homogenization steady effective equation} is challenging. However, investigating the long-term behavior of the reaction-diffusion equation is possible, as it typically converges to its similarity solution (\cite{gupta2019diffusion, wang2013self, barenblatt1996scaling, eggers2008role}). For the first type of initial condition, where the solution vanishes at infinity, similar to the classical diffusion equation, the scaling relation of equations \eqref{eq:homogenization unsteady effective equation} and \eqref{eq:homogenization steady effective equation} allows for a self-similar solution of the following form
\begin{equation}\label{eq:Self-similar solution for diffusion equation}
\begin{aligned}
c_{i,0}(x,t)= \frac{1}{\sqrt{t}} C_{i} \left( \xi \right),\quad  \xi=\frac{x}{\sqrt{t}}.
\end{aligned}
\end{equation}
The conservation of mass imposes an additional condition $\int\limits_{-\infty}^{\infty} C_{i} (\xi) \mathrm{d} \xi =\int\limits_{-\infty}^{\infty} c_{I,i} (x) \mathrm{d}x$.  With the change of variable $\tau=\log t$, $\xi=t^{-\frac{1}{2}}x$ and $\mathbf{c} (x,t) = t^{-\frac{1}{2}}\mathbf{C} (\xi,\tau)$, equation \eqref{eq:homogenization steady effective equation}  becomes
\begin{equation}\label{eq:homogenization steady effective equation similarity variables0}
\begin{aligned}
\partial_{\tau}\mathbf{C}&= \frac{1}{2}\mathbf{C}+ \frac{\xi}{2}\partial_{\xi}\mathbf{C}-\mathrm{Pe}^{2}  \partial_{\xi}\left\langle u   \Delta_{\mathbf{y}}^{-1}u \right\rangle_{\mathbf{y},\tau} \mathbf{D} (\mathbf{C})^{-1}\partial_{\xi}\mathbf{C} + \partial_{\xi} \mathbf{D} (\mathbf{C}) \partial_{\xi}\mathbf{C},
\end{aligned}
\end{equation}
where $\mathbf{C}= (C_{1},\hdots,C_{n-1})$. The self-similarity solution is the steady solution of this equation, which satisfies
\begin{equation}\label{eq:homogenization steady effective equation similarity variables}
\begin{aligned}
 0&= \frac{1}{2}\mathbf{C}+ \frac{\xi}{2}\partial_{\xi}\mathbf{C}-\mathrm{Pe}^{2}  \partial_{\xi}\left\langle u   \Delta_{\mathbf{y}}^{-1}u \right\rangle_{\mathbf{y},\tau} \mathbf{D} (\mathbf{C})^{-1}\partial_{\xi}\mathbf{C} + \partial_{\xi} \mathbf{D} (\mathbf{C}) \partial_{\xi}\mathbf{C}.
\end{aligned}
\end{equation}
 Integrating both side of the equation and using the vanishing condition at infinity reduces the equation to
\begin{equation}\label{eq:homogenization steady effective equation similarity variables order 1}
\begin{aligned}
 0&=  \frac{\xi}{2}\mathbf{C}-\mathrm{Pe}^{2}  \left\langle u   \Delta_{\mathbf{y}}^{-1}u \right\rangle_{\mathbf{y},\tau} \mathbf{D} (\mathbf{C})^{-1}\partial_{\xi}\mathbf{C} +  \mathbf{D} (\mathbf{C}) \partial_{\xi}\mathbf{C}.
\end{aligned}
\end{equation}

While the self-similarity solution of the ion concentration may not be a Gaussian distribution function, it has the property
\begin{equation}
 \int\limits_{-\infty}^{\infty} x^{n}c_{i,0} (x,t)\mathrm{d} x=  \int\limits_{-\infty}^{\infty} \frac{x^{n}}{\sqrt{t}} C_{i} \left( \frac{x}{\sqrt{t}} \right) \mathrm{d} x = t^{\frac{n}{2}}\int\limits_{-\infty}^{\infty} \xi^{n}C_{i} (\xi) \mathrm{d} \xi.
\end{equation}
 This equation implies that the second moment of the ion concentration, $\frac{1}{|\Omega|}\int\limits_{-\infty}^{\infty} x^{2}c_{i} (x,\mathbf{y},t)\mathrm{d} \mathbf{y}\mathrm{d} x$, grows linearly asymptotically for large $t$. Since $\bar{c}_{i}$ converges to $c_{i,0}$ at long times, the longitudinal effective diffusivity of $i$-th ion defined in equation \eqref{eq:effectiveDiffusivityDefinition} can be expressed in terms of $C_{i}$
\begin{equation}\label{eq:diffusivity by similarity solution}
\begin{aligned}
  \kappa_{\mathrm{eff},i}&=\lim\limits_{t\rightarrow \infty} \frac{\int\limits_{-\infty}^{\infty} x^{2}c_{i,0} (x,t)\mathrm{d} x- \left( \int\limits_{-\infty}^{\infty} xc_{i,0} (x,t)\mathrm{d} x \right)^{2}}{2t \int\limits_{-\infty}^{\infty} c_{i,0} \mathrm{d} x}=  \frac{\int\limits_{-\infty}^{\infty} \xi^{2}C_{i} (\xi) \mathrm{d} \xi- \left(  \int\limits_{-\infty}^{\infty} \xi C_{i} (\xi) \mathrm{d} \xi \right)^{2}}{2\int\limits_{-\infty}^{\infty} C_{i} (\xi)\mathrm{d} \xi}.  \\
\end{aligned}
\end{equation}
The previous definition \eqref{eq:effectiveDiffusivityDefinition} required advancing the solution of the governing equation in the full domain (a high-dimensional space) until the diffusion time scale was resolved. In contrast, the definition \eqref{eq:diffusivity by similarity solution}  present here only requires solving the steady-state solution of the effective equation that depends on one variable, which is more computationally efficient.

Due to the structure of equation \eqref{eq:homogenization steady effective equation similarity variables order 1}, we can derive an approximation of the effective as follows. Assuming $\mathbf{C}_{0}= (C_{0,1},\hdots,C_{0,n})$ and $\mathbf{C}_{\mathrm{Pe}}= (C_{\mathrm{Pe},1},\hdots,C_{\mathrm{Pe},n})$ satisfies equation, respectively,
\begin{equation}
\begin{aligned}
 &0=  \frac{\xi}{2}\mathbf{C}_{0}+  \mathbf{D} (\mathbf{C}_{0}) \partial_{\xi}\mathbf{C}_{0}, \quad
   0=  \frac{\xi}{2}\mathbf{C}_{\mathrm{Pe}}- \left\langle u   \Delta_{\mathbf{y}}^{-1}u \right\rangle_{\mathbf{y},\tau} \mathbf{D} (\mathbf{C}_{\mathrm{Pe}})^{-1}\partial_{\xi}\mathbf{C}_{\mathrm{Pe}}.
\end{aligned}
\end{equation}
Then we have an approximation for the effective diffusivity which is valid at small and large $\mathrm{Pe}$,
\begin{equation}\label{eq:approximation effective diffusivity}
\begin{aligned}
&  \kappa_{\mathrm{eff},i}\approx \frac{\int\limits_{-\infty}^{\infty} \xi^{2}C_{0,i} (\xi) \mathrm{d} \xi- \left(  \int\limits_{-\infty}^{\infty} \xi C_{0,i} (\xi) \mathrm{d} \xi \right)^{2}}{2\int\limits_{-\infty}^{\infty} C_{0,i} (\xi)\mathrm{d} \xi}+ \mathrm{Pe}^{2} \frac{\int\limits_{-\infty}^{\infty} \xi^{2}C_{\mathrm{Pe},i} (\xi) \mathrm{d} \xi- \left(  \int\limits_{-\infty}^{\infty} \xi C_{\mathrm{Pe},i} (\xi) \mathrm{d} \xi \right)^{2}}{2\int\limits_{-\infty}^{\infty} C_{\mathrm{Pe},i} (\xi)\mathrm{d} \xi}.
\end{aligned}
\end{equation}
In the numerical simulation presented in the next section, we observe that this approximation agrees with the effective diffusivity for most $\mathrm{Pe}$ values, deviating only for moderate $\mathrm{Pe}$ values (around 1 to 10).

For the second type of initial condition, where the background ion concentration is nonzero, one can search for the asymptotic expansion of the concentration field in the following form 
\begin{equation}\label{eq:Self-similar solution for diffusion equation nonzero}
\begin{aligned}
c_{i,0}(x,t)=c_{i} (\infty)+ \frac{1}{\sqrt{t}} C_{i} \left( \xi \right)+o \left( t^{-\frac{1}{2}} \right).
\end{aligned}
\end{equation}
When $c_i(\infty) \neq 0$ for $i = 1, \dots, n-1$, it is possible to simplify the nonlinear effective equation to a linear diffusion equation as time approaches infinity. In order to achieve this, we substitute this expression into equation \eqref{eq:homogenization steady effective equation} and take the limit as $t$ tends to infinity, which yields the equation for $C_i$
\begin{equation}\label{eq:homogenization steady effective equation similarity variables Non}
\begin{aligned}
 0&= \frac{1}{2}\mathbf{C}+ \frac{\xi}{2}\partial_{\xi}\mathbf{C}+ \partial_{\xi} \left(\tilde{\mathbf{D}}- \mathrm{Pe}^{2} \left\langle u   \Delta_{\mathbf{y}}^{-1}u \right\rangle_{\mathbf{y},\tau} \tilde{\mathbf{D}}^{-1} \right)\partial_{\xi}\mathbf{C} .
\end{aligned}
\end{equation}
where  $\tilde{\mathbf{D}}$ and $\tilde{\mathbf{D}}^{-1}$  are constant matrices
\begin{equation}
\begin{aligned}
&\tilde{\mathbf{D}}=
  \begin{bmatrix}
    \kappa_{1}&\hdots&0\\
    &\hdots&\\
    0&\hdots&\kappa_{n-1}
  \end{bmatrix}
-\begin{bmatrix}
  \kappa_{1}z_{1}c_{1} (\infty)\\
  \hdots\\
  \kappa_{n-1}z_{n-1}c_{n-1} (\infty)
\end{bmatrix} \frac{
  \begin{bmatrix}
(\kappa_{1}-\kappa_{n})z_{1},\hdots, (\kappa_{n-1}-\kappa_{n})z_{n-1}
  \end{bmatrix}
}{\sum\limits_{i=1}^{n-1} (z_{i} \kappa_{i}-z_{n}\kappa_{n}) z_{i}c_{i} (\infty)}, \\
&\tilde{\mathbf{D}}^{-1}=  \begin{bmatrix}
  \frac{1}{\kappa_{1}} &\hdots&0\\
    &\hdots&\\
    0&\hdots&\frac{1}{\kappa_{n-1}}
  \end{bmatrix} \left( 
    I+
    \begin{bmatrix}
  \kappa_{1}z_{1}c_{1} (\infty)\\
  \hdots\\
  \kappa_{n-1}z_{n-1}c_{n-1} (\infty)
\end{bmatrix} \frac{
  \begin{bmatrix}
\frac{(\kappa_{1}-\kappa_{n})z_{1}}{\kappa_{1}},\hdots, \frac{(\kappa_{n-1}-\kappa_{n})z_{n-1}}{\kappa_{n-1}}
  \end{bmatrix} 
}{\kappa_{n}\sum\limits_{i=1}^{n-1}  (z_{i}-z_{n}) z_{i}c_{i} (\infty)} \right).
\end{aligned}
\end{equation}
The constant diffusion tensor implies that for a non-zero background ion concentration, the perturbed concentrations satisfy a multi-dimensional diffusion equation at long times. The expression of the diffusion tensor provides a formula for measuring the mutual diffusion coefficients.  It is worth noting that if the background ion concentration is smaller compared to the perturbed concentration, the system will take a longer time to reach this long-time asymptotic state.

For the third type of initial condition, the self-similar solution takes the  form
\begin{equation}
\begin{aligned}
c_{i,0}(x,t)= C_{i} \left( \xi \right),\quad  \xi=\frac{x}{\sqrt{t}},
\end{aligned}
\end{equation}
where $C_{i} (\xi)$ solves
\begin{equation}
\begin{aligned}
0&= \frac{\xi}{2}\partial_{\xi}\mathbf{C}-\mathrm{Pe}^{2}  \partial_{\xi}\left\langle u   \Delta_{\mathbf{y}}^{-1}u \right\rangle_{\mathbf{y},\tau} \mathbf{D}^{-1}\partial_{\xi}\mathbf{C} + \partial_{\xi} \mathbf{D} (\mathbf{C}) \partial_{\xi}\mathbf{C}.
\end{aligned}
\end{equation}
It is easy to show that the second moment of the derivative of the solution grows linearly in $t$,
\begin{equation}
\begin{aligned}
\frac{1}{|\Omega|}\int\limits_{-\infty}^{\infty} x^{2}\partial_{x}c_{i,0} (x,\mathbf{y},t)\mathrm{d} \mathbf{y}\mathrm{d} x \approx \int\limits_{-\infty}^{\infty}\frac{ x^{2} }{\sqrt{t}}\partial_{\xi}C_{i} \left( \frac{x}{\sqrt{t}} \right)\mathrm{d}x= t \int\limits_{-\infty}^{\infty}\xi^{2} \partial_{\xi}C_{i} (\xi)\mathrm{d} \xi.
\end{aligned}
\end{equation}
Therefore, we can also define the effective diffusivity via the self-similarity solution,
\begin{equation}\label{eq:diffusivity by similarity solution third type}
\begin{aligned}
  \kappa_{\mathrm{eff},i}&=  \frac{\int\limits_{-\infty}^{\infty} \xi^{2}\partial_{\xi}C_{i} (\xi) \mathrm{d} \xi- \left(  \int\limits_{-\infty}^{\infty} \xi \partial_{\xi}C_{i} (\xi) \mathrm{d} \xi \right)^{2}}{2\int\limits_{-\infty}^{\infty} \partial_{\xi}C_{i} (\xi)\mathrm{d} \xi}.  \\
\end{aligned}
\end{equation}

When the diffusion tensor is constant such as the case that diffusion-induced electric potential is negligible, the first and third types of initial conditions result in the same effective diffusivity, as the equation of the self-similarity solution commute with the differential operator. However, if the diffusion tensor varies with concentration, these two types of initial conditions can yield different effective diffusivities. Nonetheless, in the examples presented in the following sections, the relative difference is less than 0.03.

\subsection{Comparison to the Taylor dispersion and reciprocal property}
\label{sec:Comparison to the Taylor dispersion}
When  the diffusion-induced electric potential is negligible, all ions are passively advected by the fluid flow.   As a result, the governing equation can be simplified to the advection-diffusion equation
\begin{equation}\label{eq:Advection Diffusion}
\begin{aligned}
\partial_{t}c_{i}+u (\mathbf{y},t)\partial_{x} c_{i}=\kappa_{i} \Delta c_{i},\quad i=1,\hdots,n.
\end{aligned}
\end{equation}
The corresponding effective equation has been reported in many literature of Taylor dispersion ( \cite{taylor2012random,ding2021enhanced,young1991shear})
\begin{equation}\label{eq:Advection Diffusion effective equation}
\begin{aligned}
\partial_{t}c_{i,0}=\left( \kappa_{i}+ \frac{\mathrm{Pe}^{2}   \left\langle u (\partial_{\tau}-\Delta)^{-1}u \right\rangle_{\mathbf{y},\tau} }{\kappa_{i}}\right) \partial_{x}^{2} c_{i,0} , \quad i=1,\hdots,n.
\end{aligned}
\end{equation}
Therefore, equation \eqref{eq:homogenization steady effective equation} can be considered to be a generalization of equation \eqref{eq:Advection Diffusion effective equation} with a nonlinear diffusion tensor taking the place of a scalar diffusion coefficient.  Additionally, both equations exhibit a ``reciprocal property'' whereby, under strong shear flow, the system behaves as if it were a different system with distinct parameters and weak flow.

To see that, using the change of variable $\kappa_{i}= \tilde{\kappa}_{i}^{-1}$, $\mathrm{Pe}= \left( \left\langle u (\partial_{\tau}-\Delta)^{-1}u \right\rangle_{\mathbf{y},\tau}  \widetilde{\mathrm{Pe}} \right)^{-1}$ and $x= \mathrm{Pe} \tilde{x} \sqrt{\left\langle u (\partial_{\tau}-\Delta)^{-1}u \right\rangle_{\mathbf{y},\tau} } $,  equation \eqref{eq:Advection Diffusion effective equation}  becomes  
\begin{equation}\label{eq:Advection Diffusion effective equation v2}
\begin{aligned}
\partial_{t}c_{i,0}= \left( \frac{ \widetilde{\mathrm{Pe}}^{2}  \left\langle u (\partial_{\tau}-\Delta)^{-1}u \right\rangle_{\mathbf{y},\tau}}{\tilde{\kappa}_{i}   }  +  \tilde{\kappa}_{i}\right) \partial_{\tilde{x}}^{2} c_{i,0}, \quad i=1,\hdots,n,
\end{aligned}
\end{equation}
which retains the same form, but with the transformed  parameters.
Hence, equation \eqref{eq:Advection Diffusion effective equation} with large P\'{e}clet numbers (representing strong flow) corresponds to equation \eqref{eq:Advection Diffusion effective equation v2} with  small P\'{e}clet numbers (representing weak flow). 

Next, we show that the effective equation \eqref{eq:homogenization steady effective equation} for the Nernst-Planck system with the steady flow  has the same property. The equivalent form of equation \eqref{eq:homogenization steady effective equation} is
\begin{equation}\label{eq:homogenization steady effective equation scalar version}
\begin{aligned}
  &\partial_{t}c_{i,0}=\mathrm{Pe}^{2} \left\langle u ( - \Delta_{\mathbf{y}})^{-1}u \right\rangle_{\mathbf{y},\tau}  \left( \frac{1}{\kappa_{i}} \partial_{x}^{2}  c_{i,0}+z_{i}  \partial_{x} \left( \frac{ c_{i,0} \sum\limits_{j=1}^{n-1} \frac{\kappa_{j}-\kappa_{n}}{\kappa_{j}}z_{j} \partial_{x} c_{j,0} }{
        \kappa_{n}\sum\limits_{i=1}^{n-1}  (z_{i}-z_{n}) z_{i}c_{i,0}}  \right) \right)\\
  &\hspace{1cm}+\kappa_{i} \partial_{x}^{2} c_{i,0}+ \kappa_{i} z_{i}  \partial_{x}   \left( \frac{ c_{i,0} \sum\limits_{j=1}^{n-1} (\kappa_{n}- \kappa_{j}) z_{j} \partial_{x} c_{j,0} }{\sum\limits_{j=1}^{n-1}(z_{j} \kappa_{j}-z_{n}\kappa_{n}) z_{j}c_{j,0}  }  \right)   , \; i=1,\hdots,n-1.\\
 \end{aligned}
\end{equation}
After rescaling using $\kappa_{i}= \frac{1}{\tilde{\kappa}_{i}}$, $\mathrm{Pe}= \left( \left\langle u (\partial_{\tau}-\Delta)^{-1}u \right\rangle_{\mathbf{y},\tau}  \widetilde{\mathrm{Pe}} \right)^{-1}$, $c_{i,0}=\frac{\tilde{c}_{i,0}}{\tilde{\kappa}_{i}}$,  $z_{i}=\tilde{\kappa}_{i}\tilde{z}_{i}$, and $x= \mathrm{Pe} \tilde{x} \sqrt{\left\langle u (-\Delta)^{-1}u \right\rangle_{\mathbf{y},\tau} }$, the above equation becomes
\begin{equation}\label{eq:homogenization steady effective equation scalar version v2}
\begin{aligned}
  &\partial_{t}\tilde{c}_{i,0}= \tilde{\kappa}_{i} \partial_{\tilde{x}}^{2} \tilde{c}_{i,0}+  \tilde{z}_{i}\tilde{\kappa}_{i}  \partial_{\tilde{x}} \left( \frac{ \tilde{c}_{i,0} \sum\limits_{j=1}^{n-1} (\tilde{\kappa}_{n}-\tilde{\kappa}_{j})\tilde{z}_{j} \partial_{\tilde{x}} \tilde{c}_{j,0} }{
      \sum\limits_{i=1}^{n-1}  (\tilde{z}_{i}\tilde{\kappa}_{i}-\tilde{z}_{n}\tilde{\kappa}_{n}) \tilde{z}_{i}\tilde{c}_{i,0}}  \right) \\
&\hspace{1cm}+ \widetilde{\mathrm{Pe}}^{2} \left\langle u ( - \Delta_{\mathbf{y}})^{-1}u \right\rangle_{\mathbf{y},\tau}  \left( \frac{1}{\tilde{\kappa}_{i}} \partial_{\tilde{x}}^{2}  \tilde{c}_{i,0}+\tilde{z}_{i}  \partial_{\tilde{x}} \left( \frac{ \tilde{c}_{i,0} \sum\limits_{j=1}^{n-1} \frac{\tilde{\kappa}_{j}-\tilde{\kappa}_{n}}{\tilde{\kappa}_{j}}\tilde{z}_{j} \partial_{\tilde{x}} \tilde{c}_{j,0} }{
        \tilde{\kappa}_{n}\sum\limits_{i=1}^{n-1}  (\tilde{z}_{i}-\tilde{z}_{n}) \tilde{z}_{i}\tilde{c}_{i,0}}  \right) \right) , \; i=1,\hdots,n-1.\\
 \end{aligned}
\end{equation}
Similar to the scenario where the diffusion-induced electric potential is negligible, after the change of variable, the resulting equation takes the same form as equation \eqref{eq:homogenization steady effective equation scalar version}, albeit with different parameters.    Hence, equation \eqref{eq:homogenization steady effective equation scalar version} with large P\'{e}clet numbers (representing strong flow) corresponds to equation \eqref{eq:homogenization steady effective equation scalar version v2} with  small P\'{e}clet numbers (representing weak flow).

The reciprocal property observed in the effective equations has two implications. First, in the limit of large P\'{e}clet numbers, it simplifies the problem to the Nernst-Planck equation in the absence of flow. Second, it establishes a correspondence between phenomena observed in systems with and without flow, allowing us to expect similar behavior in different systems.

Lastly, akin to the governing equation \eqref{eq:Advection-Nernst-Planck equation no potential}, the effective equations also demonstrate the following scaling properties. This property states that any solution multiplied by a constant remains a valid solution to the system. Moreover, if all valences are multiplied by a constant, the original solution of the system remains valid for the system with the new valences

\section{Theoretical results for two or three different ion species}
\label{sec:Analytical result}
In this section, a series of examples will be analyzed to gain a deeper understanding of how individual ion diffusivities interact and impact the overall dynamics of dissolved salt.

\subsection{Two different ion species}
We first consider the simplest example where the solution consists of two different type of ion species. When $n=2$,  the diffusion tensor provided  in equation \eqref{eq:multiscale expansion epsilon -1 simplified} and its inverse matrix are scalars.  Effective equation \eqref{eq:homogenization steady effective equation} becomes a diffusion equation
\begin{equation}\label{eq:homogenization steady effective equation two ions}
\begin{aligned}
 &  \partial_{t}c_{1,0}= \kappa_{\mathrm{eff},1}  \partial_{x}^{2}c_{1,0},\quad \kappa_{\mathrm{eff},1}=\left(\frac{\kappa _1 \kappa _2 \left(z_1-z_2\right)}{\kappa _1 z_1-\kappa _2 z_2} - \mathrm{Pe}^{2}\left\langle u  \Delta_{\mathbf{y}}^{-1}u \right\rangle_{\mathbf{y},\tau}\frac{\kappa _1 z_1-\kappa _2 z_2}{\kappa _1 \kappa _2 \left(z_1-z_2\right)}        \right).
\end{aligned}
\end{equation}
\cite{deen1998analysis} shows that, in absence of flow, the Nernst-Planck equation reduces to a diffusion equation with a constant diffusion coefficient $\frac{\kappa _1 \kappa _2 \left(z_1-z_2\right)}{\kappa _1 z_1-\kappa _2 z_2}$. The calculation here verifies that this result also holds in presence of the shear flow. Therefore, the transport of binary electrolyte solution can be described by the classical Taylor dispersion theory.

\subsection{Three different ion species}
Many physical systems contains three different ion species, such as  the ternary electrolyte solutions and the mixture of two the binary electrolyte solutions, {\color{blue}e.g.,} the mixture of  sodium fluorescein and sodium chloride. When $n=3$, the diffusion tensor provided  in equation \eqref{eq:multiscale expansion epsilon -1 simplified} and its inverse matrix depend on the ion concentrations, in contrast to the case with $n=2$,
 \begin{equation}\label{eq:diffussion tensor three ions}
\begin{aligned}
  &\mathbf{D}=
  \begin{bmatrix}
   \kappa _1-\frac{c_1 \kappa _1 \left(\kappa _1-\kappa _3\right) z_1^2}{c_1 z_1 \left(\kappa _1 z_1-\kappa _3 z_3\right)+c_2 z_2 \left(\kappa _2 z_2-\kappa _3 z_3\right)} & -\frac{c_1 \kappa _1 \left(\kappa _2-\kappa _3\right) z_1 z_2}{c_1 z_1 \left(\kappa _1 z_1-\kappa _3 z_3\right)+c_2 z_2 \left(\kappa _2 z_2-\kappa _3 z_3\right)} \\
 -\frac{c_2 \kappa _2 \left(\kappa _1-\kappa _3\right) z_1 z_2}{c_1 z_1 \left(\kappa _1 z_1-\kappa _3 z_3\right)+c_2 z_2 \left(\kappa _2 z_2-\kappa _3 z_3\right)} & \kappa _2-\frac{c_2 \kappa _2 \left(\kappa _2-\kappa _3\right) z_2^2}{c_1 z_1 \left(\kappa _1 z_1-\kappa _3 z_3\right)+c_2 z_2 \left(\kappa _2 z_2-\kappa _3 z_3\right)} \\  
  \end{bmatrix},
\\
&\mathbf{D}^{-1}=
\begin{bmatrix}
   \frac{c_2 \kappa _3 z_2 \left(z_2-z_3\right)+c_1 z_1 \left(\kappa _1 z_1-\kappa _3 z_3\right)}{\kappa _1 \kappa _3 \left(c_1 z_1 \left(z_1-z_3\right)+c_2 z_2 \left(z_2-z_3\right)\right)} & \frac{c_1 \left(\kappa _2-\kappa _3\right) z_1 z_2}{\kappa _2 \kappa _3 \left(c_1 z_1 \left(z_1-z_3\right)+c_2 z_2 \left(z_2-z_3\right)\right)} \\
 \frac{c_2 \left(\kappa _1-\kappa _3\right) z_1 z_2}{\kappa _1 \kappa _3 \left(c_1 z_1 \left(z_1-z_3\right)+c_2 z_2 \left(z_2-z_3\right)\right)} & \frac{c_1 \kappa _3 z_1 \left(z_1-z_3\right)+c_2 z_2 \left(\kappa _2 z_2-\kappa _3 z_3\right)}{\kappa _2 \kappa _3 \left(c_1 z_1 \left(z_1-z_3\right)+c_2 z_2 \left(z_2-z_3\right)\right)} \\
\end{bmatrix}.
\end{aligned}
\end{equation}

\subsubsection{Exact solutions}
In contrast to the binary electrolyte case, the presence of nonlinearity in the system makes it generally challenging to obtain an exact self-similarity solution. However, in certain special cases, we can still derive exact solutions. For the sake of simplicity and without loss of generality, let us assume that the first and second ion species carry charges of the same sign, whereas the third ion species carries a charge of the opposite sign.

The first special case arises when $\kappa_{1}=\kappa_{2}$. In this situation, we can effectively treat the first and second types of ions as a single type, thereby simplifying the system to a scenario with two ion species. For the first type initial condition, the equation \eqref{eq:homogenization steady effective equation similarity variables order 1}   for the self similarity solution becomes
\begin{equation}
\begin{aligned}
  &0=\frac{\xi}{2}C_{1}+ \sigma \partial_{\xi}C_{1}= \frac{\xi}{2}C_{2}+ \sigma \partial_{\xi}C_{2},\\
  &\sigma=\left( \frac{\kappa _1 \left(\kappa _1 z_2 \left(z_2-z_1\right)+\kappa _3 \left(z_1+z_2\right) \left(z_1-z_3\right)\right)}{\kappa _1 \left(z_1^2+z_2^2\right)-\kappa _3 \left(z_1+z_2\right) z_3}  \right.\\
  &\hspace{0.5cm}\left. +\frac{\left\langle u  (- \Delta_{\mathbf{y}})^{-1}u \right\rangle_{\mathbf{y},\tau}   \mathrm{Pe}^2 \left(\kappa _1 z_1 \left(z_1+z_2\right)-\kappa _3 \left(z_2 \left(z_3-z_2\right)+z_1 \left(z_2+z_3\right)\right)\right)}{\kappa _1 \kappa _3 \left(z_1^2+z_2^2-\left(z_1+z_2\right) z_3\right)}\right).
\end{aligned}
\end{equation}
The self similarity solutions are given by 
\begin{equation}\label{eq:exact solution same diffusivity}
\begin{aligned}
&C_{1} (\xi) =  \frac{\int\limits_{-\infty}^{\infty}c_{I,1} (x)\mathrm{d}x }{\sigma\sqrt{2\pi}} e^{-\frac{1}{2} \left( \frac{\xi}{\sigma} \right)^{2}},\quad C_{2} (\xi) =  \frac{\int\limits_{-\infty}^{\infty}c_{I,2} (x)\mathrm{d}x }{\sigma\sqrt{2\pi}} e^{-\frac{1}{2} \left( \frac{\xi}{\sigma} \right)^{2}}.
\end{aligned}
\end{equation}

The second special case arises when $\kappa_{2}=\kappa_{3}$, where the diffusion tensor \eqref{eq:diffussion tensor three ions} becomes
\begin{equation}
\begin{aligned}
&\mathbf{D}=
\begin{bmatrix}
 \kappa _1 & -\frac{\kappa _1 \left(\kappa _2-\kappa _1\right) z_1 z_2 c_1}{z_1 c_1 \left(\kappa _1 z_1-\kappa _1 z_3\right)+z_2 c_2 \left(\kappa _2 z_2-\kappa _1 z_3\right)} \\
 0 & \kappa _2-\frac{\kappa _2 \left(\kappa _2-\kappa _1\right) z_2^2 c_2}{z_1 c_1 \left(\kappa _1 z_1-\kappa _1 z_3\right)+z_2 c_2 \left(\kappa _2 z_2-\kappa _1 z_3\right)} \\
\end{bmatrix},  \\
&\mathbf{D}^{-1}=
\begin{bmatrix}
 \frac{\kappa _1 z_1^2 c_1-\kappa _1 z_3 z_1 c_1+\kappa _1 z_2 \left(z_2-z_3\right) c_2}{\kappa _1^2 \left(z_1^2 c_1-z_3 z_1 c_1+z_2 \left(z_2-z_3\right) c_2\right)} & \frac{\left(\kappa _2-\kappa _1\right) z_1 z_2 c_1}{\kappa _1 \kappa _2 \left(z_1^2 c_1-z_3 z_1 c_1+z_2 \left(z_2-z_3\right) c_2\right)} \\
 0 & \frac{\kappa _1 z_1^2 c_1-\kappa _1 z_3 z_1 c_1+z_2 c_2 \left(\kappa _2 z_2-\kappa _1 z_3\right)}{\kappa _1 \kappa _2 \left(z_1^2 c_1-z_3 z_1 c_1+z_2 \left(z_2-z_3\right) c_2\right)} \\
\end{bmatrix}. \\
\end{aligned}
\end{equation}
The self similarity solutions are
\begin{equation}\label{eq:exact solution non zero Pe ex1}
\begin{aligned}
  &C_{1}= a_{2} g (\xi)^{\frac{d_2 \kappa _1}{d_3 \left(\kappa _2-\kappa _1\right)}} h (\xi),\quad C_{2}=a_{2}g (\xi)^{\frac{d_1 \kappa _2}{d_3 \left(\kappa _2-\kappa _1\right)}}  h (\xi),  \\
  &h (\xi) =\frac{\left(z_{1}+z_{2}g (\xi) \right)^{\frac{\kappa _1 \left(z_2-z_1\right) z_2 \left(\kappa _3^2+\mathrm{Pe}^2\right)}{d_1 \left(\kappa _1 \left(z_1-z_2\right)+\kappa _3 \left(z_2-z_3\right)\right)}}}
  {\left( \kappa _1 z_1 \left(z_1-z_3\right)d_{3}-z_2 d_{4}  g (\xi) \right)^{\frac{ -\left\langle u   \Delta_{\mathbf{y}}^{-1}u \right\rangle_{\mathbf{y},\tau}\mathrm{Pe}^2 d_1 \kappa _1 \left(\kappa _1-\kappa _2\right) \kappa _2^2  z_2^2}{d_3 d_5 d_4}} },\\
   &d_1= \kappa _1^2-\left\langle u   \Delta_{\mathbf{y}}^{-1}u \right\rangle_{\mathbf{y},\tau}\mathrm{Pe}^2,\quad d_2= \kappa _2^2-\left\langle u   \Delta_{\mathbf{y}}^{-1}u \right\rangle_{\mathbf{y},\tau}\mathrm{Pe}^2, \\
  &d_3= -\left\langle u   \Delta_{\mathbf{y}}^{-1}u \right\rangle_{\mathbf{y},\tau}\mathrm{Pe}^2-\kappa _1 \kappa _2,\\
  &d_4= -\left\langle u   \Delta_{\mathbf{y}}^{-1}u \right\rangle_{\mathbf{y},\tau}\mathrm{Pe}^2 \left(\kappa _1 z_3-\kappa _2 z_2\right)+\kappa _1^2 \kappa _2 \left(z_2-z_3\right),\\
  &d_5= -\left\langle u   \Delta_{\mathbf{y}}^{-1}u \right\rangle_{\mathbf{y},\tau}\mathrm{Pe}^2 \left(\kappa _2 z_2-\kappa _1 z_1\right)+\kappa _1^2 \kappa _2 \left(z_1-z_2\right).
 \end{aligned}
\end{equation}
where $g (\xi)= f^{-1} \left( a_{1}-\frac{\xi ^2 \kappa _1 \left(\kappa _1-\kappa _2\right) \left(z_1-z_3\right)}{4 \left(\kappa _1^2-\left\langle u   \Delta_{\mathbf{y}}^{-1}u \right\rangle_{\mathbf{y},\tau}\mathrm{Pe}^2\right)} \right) $. $a_{1}$ and $a_{2}$ are constants that can be determined by the total mass of each ion species. $f ^{-1}$  is the inverse of the following function
\begin{equation}
\begin{aligned}
  f(x)&=\frac{\left\langle u   \Delta_{\mathbf{y}}^{-1}u \right\rangle_{\mathbf{y},\tau} \mathrm{Pe}^2 d_1 \kappa _1 \left(\kappa _1-\kappa _2\right){}^2 \kappa _2^2  z_2^2 \left(z_1-z_3\right) \log \left( d_4 z_2 x+d_3 \kappa _1 \left(z_3-z_1\right) z_1\right)}{d_3 d_4 d_5}\\
  &  -\frac{\log \left(x z_2+z_1\right) \left(-\left\langle u   \Delta_{\mathbf{y}}^{-1}u \right\rangle_{\mathbf{y},\tau}\mathrm{Pe}^2 \left(\kappa _1 z_1  -\kappa _2 z_2\right){}^2+\kappa _1^2 \kappa _2^2 \left(z_1-z_2\right){}^2\right)}{d_5}\\
 &+\frac{d_2 \kappa _1 \left(z_3-z_1\right) \log (x)}{d_3}.
\end{aligned}
\end{equation}
In general, the close form expression of the moment of the above exact solution is unavailable, necessitating the computation of the effective diffusivity through numerical integration.

Next, we present two examples that will be discussed in the following section.  When the diffusivities and valences are  $\kappa_{1}=1, \kappa_{2}=0.1, \kappa_{3}=1, z_{1}=1, z_{2}=1, z_{3}=-2$, and $\mathrm{Pe}=0$, the self similarity solutions are
\begin{equation}\label{eq:exact solution ex1}
\begin{aligned}
&C_{1} (\xi) =\frac{a_2 e^{\frac{a_1}{27}-\frac{\xi ^2}{4}}}{\sqrt[3]{e^{\frac{a_1}{3}-\frac{9 \xi ^2}{4}}+1}},\quad C_{2} (\xi) =\frac{a_2 e^{\frac{10 a_1}{27}-\frac{5 \xi ^2}{2}}}{\sqrt[3]{e^{\frac{a_1}{3}-\frac{9 \xi ^2}{4}}+1}}. \\
\end{aligned}
\end{equation}
If $\int\limits_{-\infty}^{\infty}C_{1} (\xi)\mathrm{d}\xi=\int\limits_{-\infty}^{\infty}C_{2} (\xi)\mathrm{d}\xi=1$, then we have  $a_{1}\approx 4.18439, a_{2}\approx 0.29164$ and $\kappa_{\mathrm{eff,}1}\approx 1.1774$, $\kappa_{\mathrm{eff},2} \approx 0.117514$, $\kappa_{\mathrm{eff},3} \approx 0.647457$.

When the diffusivities and valences are $\kappa_{1}=1, \kappa_{2}=10, \kappa_{3}=1, z_{1}=2, z_{2}=2, z_{3}=-3$ and $\mathrm{Pe}=0$, the self similarity solutions are
\begin{equation}\label{eq:exact solution ex2}
\begin{aligned}
&C_{1} (\xi) =\frac{a_2 e^{-\frac{2 a_1}{9}-\frac{\xi ^2}{4}}}{\left(2 e^{\frac{1}{5} \left(a_1+\frac{9 \xi ^2}{8}\right)}+2\right){}^{2/5}}, \quad C_{2} (\xi) =\frac{a_2 e^{-\frac{8 a_1+9 \xi ^2}{360} }}{\left(2 e^{\frac{1}{5} \left(a_1+\frac{9 \xi ^2}{8}\right)}+2\right){}^{2/5}}. \\
\end{aligned}
\end{equation}
If $\int\limits_{-\infty}^{\infty}C_{1} (\xi)\mathrm{d}\xi=\int\limits_{-\infty}^{\infty}C_{2} (\xi)\mathrm{d}\xi=1$, then we have  $a_{1}\approx -3.12729, a_{2}\approx 0.23872$ and $\kappa_{\mathrm{eff,}1}\approx 0.825144$, $\kappa_{\mathrm{eff},2} \approx 2.55403$, $\kappa_{\mathrm{eff},3}\approx 1.84843$.

Last, for certain combinations of valences, the exact solution is obtainable. However, the solution may become lengthy when the P\'eclet number is nonzero. Here, we only present the exact solutions for the combination where $z_{1}=1, z_{2}=1, z_{3}=-1$, and $\mathrm{Pe}=0$, which has been used in many studies (\cite{gupta2019diffusion,rodrigo2022ternary,ribeiro2019coupled,price1988theory}),
\begin{equation}\label{eq:exact solution ex3}
\begin{aligned}
&C_1(\xi )= \frac{a_2 e^{-\frac{\xi ^2 \left(\kappa _1+\kappa _3\right)}{8 \kappa _1 \kappa _3}}}{\sqrt{a_1 e^{\frac{1}{4} \xi ^2 \left(\frac{1}{\kappa _1}-\frac{1}{\kappa _2}\right)}+1}}, \quad C_2(\xi )= \frac{\sqrt{a_1} a_2 e^{-\frac{1}{8} \xi ^2 \left(\frac{2}{\kappa _2}+\frac{1}{\kappa _3}-\frac{1}{\kappa _1}\right)}}{\sqrt{ e^{\frac{1}{4} \xi ^2 \left(\frac{1}{\kappa _1}-\frac{1}{\kappa _2}\right)}+1}}. \\
\end{aligned}
\end{equation}

\subsubsection{Large diffusivity discrepancy}

In numerous scenarios, there is often a significant disparity in diffusivity between ions, with one ion being extremely diffusive compared to the others, or conversely, one ion exhibiting significantly slower diffusion. For example, this discrepancy is frequently observed in systems under acidic conditions, where hydrogen ions can be nearly 10 times faster than all other ions (\cite{vanysek1993ionic}). Conversely, in the presence of larger ions, such as polyelectrolytes or buffered proteins (\cite{leaist1993diffusion}), their diffusivity may be smaller compared to the other ions.

Therefore, it is intriguing to examine the dynamics of the system when there is a significant difference in diffusivity between ions. First, we consider the limit of large diffusivity. One might anticipate that if the diffusivity of one ion species tends to infinity, the effective diffusivity of all ion species would diverge. However, the asymptotic expansion of equation \eqref{eq:exact solution ex3} for $\kappa_{2}=\infty$ is
\begin{equation}
\begin{aligned}
  &C_1(\xi )=a_2 \left( \frac{a_1  \xi ^2 e^{\frac{\xi ^2}{4 \kappa _1}-\frac{\xi ^2 \left(\kappa _1+\kappa _3\right)}{8 \kappa _1 \kappa _3}}}{8 \kappa _2 \left(a_1 e^{\frac{\xi ^2}{4 \kappa _1}}+1\right){}^{3/2}}+\frac{ e^{-\frac{\xi ^2 \left(\kappa _1+\kappa _3\right)}{8 \kappa _1 \kappa _3}}}{\sqrt{a_1 e^{\frac{\xi ^2}{4 \kappa _1}}+1}}+\mathcal{O}\left(\kappa _2^{-2}\right) \right),\\
 &C_2(\xi )=a_2  \left( \frac{a_1 e^{\frac{1}{8} \xi ^2 \left(\frac{1}{\kappa _1}-\frac{1}{\kappa _3}\right)}}{\sqrt{c_1 e^{\frac{\xi ^2}{4 \kappa _1}}+1}}-\frac{a_1 \xi ^2 e^{\frac{1}{8} \xi ^2 \left(\frac{1}{\kappa _1}-\frac{1}{\kappa _3}\right)} \left(a_1 e^{\frac{\xi ^2}{4 \kappa _1}}+2\right)}{8 \kappa _2 \left(a_1 e^{\frac{\xi ^2}{4 \kappa _1}}+1\right){}^{3/2}}+\mathcal{O}\left(\kappa _2^{-2}\right) \right),
\end{aligned}
\end{equation}
which reveals that, contrary to expectations, the self-similarity solution converges to a limiting distribution, and the effective diffusivities of the ion species converge to a finite value. To demonstrate this more rigorously, we consider the limit as $\kappa_{2} \rightarrow \infty$. The diffusion tensor \eqref{eq:homogenization steady effective equation similarity variables order 1} takes the following form:
\begin{equation}
\begin{aligned}
  &\mathbf{D}=
  \begin{bmatrix}
 \kappa _1 & -\frac{\kappa _1 z_1 c_1}{z_2 c_2} \\
 -\frac{\left(\kappa _1-\kappa _3\right) z_1}{z_2} & \frac{\kappa _1 z_1^2 c_1-\kappa _3 z_3 z_1 c_1+\kappa _3 z_2^2 c_2-\kappa _3 z_2 z_3 c_2}{z_2^2 c_2} \\
\end{bmatrix},\\
&\mathbf{D}^{-1}=
  \begin{bmatrix}
 \frac{\kappa _1 z_1^2 c_1-\kappa _3 z_3 z_1 c_1+\kappa _3 z_2 \left(z_2-z_3\right) c_2}{\kappa _1 \kappa _3 \left(z_1^2 c_1-z_3 z_1 c_1+z_2 \left(z_2-z_3\right) c_2\right)} & \frac{z_1 z_2 c_1}{\kappa _3 \left(z_1^2 c_1-z_3 z_1 c_1+z_2 \left(z_2-z_3\right) c_2\right)} \\
 \frac{\left(\kappa _1-\kappa _3\right) z_1 z_2 c_2}{\kappa _1 \kappa _3 \left(z_1^2 c_1-z_3 z_1 c_1+z_2 \left(z_2-z_3\right) c_2\right)} & \frac{z_2^2 c_2}{\kappa _3 \left(z_1^2 c_1-z_3 z_1 c_1+z_2 \left(z_2-z_3\right) c_2\right)} \\
\end{bmatrix}.
\end{aligned}
\end{equation}

In this case, the self similarity solutions are
\begin{equation}\label{eq:exact solution large diffusivity}
\begin{aligned}
  &C_{1}= a_{2} g (\xi)^{\frac{\kappa _3 z_3-\kappa _1 z_1}{\kappa _1 \left(z_1-z_2\right)+\kappa _3 \left(z_2-z_3\right)}} h (\xi),
  \quad C_{2}=a_{2}g (\xi)^{\frac{\left(\kappa _1-\kappa _3\right) z_2}{ -\kappa _1 z_{1} +\left(\kappa _1-\kappa _3\right) z_2+\kappa _3 z_3}}  h (\xi),  \\
  &h (\xi) =\frac{\left(z_{1}+z_{2}g (\xi) \right)^{\frac{d_1 \kappa _2 \left(z_1-z_2\right) \left(\kappa _1 z_1-\kappa _2 z_2\right)}{d_5 \left(\kappa _2-\kappa _1\right) \left(z_1-z_3\right)}}}
  {\left( \kappa _1 \kappa _3 z_1 \left(z_3-z_1\right)+d_2 z_2 g (\xi) \right)^{\frac{-d_3^2   \mathrm{Pe}^2 z_2 d_3}{d_1 d_2 \left(\kappa _1 \left(z_1-z_2\right)+\kappa _3 \left(z_2-z_3\right)\right)}} },\\
  &d_1=-\left\langle u   \Delta_{\mathbf{y}}^{-1}u \right\rangle_{\mathbf{y},\tau}\mathrm{Pe}^2 z_2+\kappa _1 \kappa _3 \left(z_1-z_2\right), \quad d_2= -\left\langle u   \Delta_{\mathbf{y}}^{-1}u \right\rangle_{\mathbf{y},\tau}\mathrm{Pe}^2 z_2+\kappa _1 \kappa _3 \left(z_3-z_2\right),\\
  &d_3= -\left\langle u   \Delta_{\mathbf{y}}^{-1}u \right\rangle_{\mathbf{y},\tau}\mathrm{Pe}^2 z_2 \left(\kappa _3 z_3-\kappa _1 z_1\right)+\kappa _1 \kappa _3 \left(\kappa _1 \left(z_2-z_1\right) z_3+\kappa _3 z_1 \left(z_3-z_2\right)\right),\\
  &d_4= \kappa _1^2 \left(z_1-z_2\right)^2-\left\langle u   \Delta_{\mathbf{y}}^{-1}u \right\rangle_{\mathbf{y},\tau}\mathrm{Pe}^2 z_2^2,\\
  &d_5=\kappa _1^2 \kappa _3^2 \left(z_1-z_3\right)^2-\left\langle u   \Delta_{\mathbf{y}}^{-1}u \right\rangle_{\mathbf{y},\tau} \mathrm{Pe}^2 \left(\kappa _1 z_1-\kappa _3 z_3\right){}^2.
 \end{aligned}
\end{equation}
where $g (\xi)= f^{-1} \left( a_{1}+\frac{1}{4} \xi ^2 \left(\kappa _1 \left(z_1-z_2\right)+\kappa _3 \left(z_2-z_3\right)\right) \right) $. $a_{1}$ and $a_{2}$ are constants that can be determined by the total mass of each ion species. $f ^{-1}$  is the inverse of the following function $f (x)$.
\begin{equation}
\begin{aligned}
  f(x)&=\frac{d_5 \log (x)}{\kappa _1 \kappa _3 \left(z_1-z_3\right)}-\frac{d_4 \left(\kappa _3^2-\left\langle u   \Delta_{\mathbf{y}}^{-1}u \right\rangle_{\mathbf{y},\tau}\mathrm{Pe}^2\right) \log \left(x z_2+z_1\right)}{d_1}\\
  &-\frac{-\left\langle u   \Delta_{\mathbf{y}}^{-1}u \right\rangle_{\mathbf{y},\tau} \mathrm{Pe}^2 d_3^2 \log \left(\kappa _1 \kappa _3 z_1 \left(z_1-z_3\right)-x d_2 z_2\right)}{d_1 d_2 \kappa _1 \kappa _3 \left(z_1-z_3\right)}.
\end{aligned}
\end{equation}

Next, we consider the same parameters used in equation \eqref{eq:exact solution ex2}, except for the diffusivity of the second ion species, which is assumed to be infinite in this case. Using the formula above, the self similarity solutions are
\begin{equation}
\begin{aligned}
&C_{1}(\xi)=\frac{a_2 e^{-\frac{\xi ^2}{4}}}{\left(a_1 e^{\frac{\xi ^2}{4}}+1\right){}^{2/5}},\quad C_2(\xi )= \frac{a_1 a_2}{\left(a_1 e^{\frac{\xi ^2}{4}}+1\right){}^{2/5}}.
\end{aligned}
\end{equation}
For $\int\limits_{-\infty}^{\infty}C_{1} (\xi)\mathrm{d}\xi=\int\limits_{-\infty}^{\infty}C_{2} (\xi)\mathrm{d}\xi=1$,  we have $a_{1}\approx 0.48018$, $a_{2}\approx 0.35928$, $\kappa_{\mathrm{eff,}1}\approx 0.647516$, $\kappa_{\mathrm{eff},2}\approx 3.04934$, and $\kappa_{\mathrm{eff},3}\approx 0.987248$. Several observations can be made from these results. First, when comparing these approximations with the effective diffusivities obtained using equation \eqref{eq:exact solution ex2}, it becomes evident that the error of this asymptotic approximation is on the order of $\kappa_2^{-1}$. To obtain a more accurate approximation, it is necessary to calculate the $\kappa_2^{-1}$ terms in the asymptotic expansion. Second, in this case, the effective diffusivities of all three ion species are lower than their respective bare diffusivities. Additionally, even if the diffusivity of the second ion species is extremely high, the diffusion-induced electric potential constrains the effective diffusivity of that particular ion species. Lastly, when the third ion species is significantly more diffusive compared to the remaining ions, the results exhibit similar behavior. However, for brevity, we will omit the discussion of this case here.

We consider the limit of small diffusivity, which can be divided into two cases. In the first case, we consider the limit $\kappa_{3}\rightarrow 0$. In the absence of flow, all self-similarity solutions collapse to a Dirac delta function, resulting in the vanishing of all effective diffusivities. To obtain nontrivial results, it is necessary to consider terms of $O(\kappa_{3})$ in the asymptotic expansion. In the presence of flow, the dominant term in the effective equation \eqref{eq:homogenization steady effective equation similarity variables order 1} is $ \mathbf{D}^{-1}$, which can be expressed in the following form
\begin{equation}
\begin{aligned}
&\mathbf{D}^{-1}=
\begin{bmatrix}
 \frac{z_1^2 c_1}{\kappa _3 d_{1}}+\frac{z_2 \left(z_2-z_3\right) c_2-z_1 z_3 c_1}{\kappa _1 d_{1}} & \frac{z_1 z_2 c_1}{\kappa _3d_{1}}-\frac{z_1 z_2 c_1}{\kappa _2d_{1}} \\
 \frac{z_1 z_2 c_2}{\kappa _3 d_{1}}-\frac{z_1 z_2 c_2}{\kappa _1 d_{1}} & \frac{z_2^2 c_2}{\kappa _3 d_{1}}+\frac{z_1^2 c_1-z_3 z_1 c_1-z_2 z_3 c_2}{\kappa _2 d_{1}} \\
\end{bmatrix}
,\\
&d_{1}=z_1^2 c_1-z_3 z_1 c_1+z_2 \left(z_2-z_3\right) c_2.
\end{aligned}
\end{equation}
Therefore, for non zero P\'eclet numbers, the effective diffusivities scale as $\kappa_{3}^{-1}$ for small $\kappa_{3}$.

In the second case, we consider the limit $\kappa_{2}\rightarrow 0$. Unlike the limit of $\kappa_{3}\rightarrow 0$, in this scenario, the concentration distribution for some ion species converges to a non-trivial limiting distribution. Taking equation \eqref{eq:exact solution ex3} as an example, for small $\kappa_{2}$, the concentration distribution for the second ion species becomes localized near $\xi=0$. The distribution of the first ion species can be approximated by a Gaussian function for $\xi$ away from the origin. The distribution of the third ion species is similar to the distribution of the second ion species near the origin, but closer to the distribution of the first ion species away from the origin.

The effective diffusivities converge to the following values:
\begin{equation}\label{eq:effective diffusivity small diffusivity}
\begin{aligned}
&\kappa_{\mathrm{eff},1}=\frac{\kappa _1 \kappa _3 \left(z_1-z_3\right)}{\kappa _1 z_1-\kappa _3 z_3}, \; \kappa_{\mathrm{eff},2}=0, \; \kappa_{\mathrm{eff},3}= \frac{z_{1} \int\limits_{-\infty}^{\infty} c_{I,1} (x)\mathrm{d} x}{ \sum\limits_{j=1}^{2}z_{j} \int\limits_{-\infty}^{\infty} c_{I,j} (x)\mathrm{d} x} \frac{\kappa _1 \kappa _3 \left(z_1-z_3\right)}{\kappa _1 z_1-\kappa _3 z_3}.
\end{aligned}
\end{equation}
We can interpret these results as follows: Some ions from the third  species bind with the second ion species and remain localized at the origin, while the remaining ions bind from the third species to the ions from the first species and diffuse throughout the channel. Consequently, the difference in diffusivities leads to ion separation. We will further investigate this phenomenon in Section \ref{sec:ion separation}.

In the presence of flow, the dominant term in the effective equation \eqref{eq:homogenization steady effective equation similarity variables order 1} is $\mathbf{D}^{-1}$, and the effective diffusivity scales as $\kappa_{2}^{-1}$ for small $\kappa_{2}$.

\section{Numerical results}
\label{sec:Examples and numerical tests}
 In this section, we will investigate the electrolyte transport through numerical simulations. The numerical simulations employed in this study hold several implications. Firstly, it is important to note that the effective equation is a valid approximation at the diffusion timescale, wherein the concentration field becomes homogenized across the channel. However, prior to reaching the diffusion timescale, the asymptotic results obtained through the homogenization method are not applicable. Therefore, we rely on numerical simulations to examine the dynamics of the concentration during this initial stage. Secondly, we utilize numerical simulations to validate the accuracy of the effective equation obtained through the homogenization method at the diffusion timescale. The results indicate that the solution of the effective equation reliably approximates the solution of the full governing equation. Thirdly, the effective equation \eqref{eq:homogenization steady effective equation}, which incorporates nonlinearity, introduces various intriguing phenomena that are not observable in binary electrolyte solutions. These phenomena will be explored through numerical simulations.

  For our simulations, we employ the Fourier spectral method as described in (\cite{ding2022determinism}), which utilizes an implicit-explicit third-order Runge-Kutta method. Specifically, we employ the explicit Runge-Kutta method to integrate the advection terms and diffusion-induced electric potential terms, while the diffusion term is integrated using the implicit Runge-Kutta method.

 The computational domain is $(x,y)\in [-8\pi, 8\pi]\times [0,1]$. The shear flow is $u (y)=\mathrm{Pe}\cos 2\pi y$ and the corresponding effective equation is provided in equation \eqref{eq:homogenization steady effective equation cos}.  We choose this flow for two reasons. First, it can be fully resolved in the Fourier spectral algorithm and ensure higher accuracy. Second, the flow profile is close to the pressure-driven flow. When the background concentration is nonzero, the system can be described by the Taylor dispersion theory.
 In this section, the initial conditions, diffusivities and valences are assumed to be of the following form unless stated otherwise,
 \begin{equation}\label{eq:numerical simulation initial conditoin}
\begin{aligned}
&c_{I,1}=c_{I,2}=\frac{e^{-\frac{1}{2} \left( \frac{x}{\sigma} \right)^{2}}}{\sigma\sqrt{2\pi}} ,  \; \sigma= \frac{1}{4},\;\kappa_{1}=1, \kappa_{2}=0.1, \kappa_{3}=1, z_{1}=1, z_{2}=1, z_{3}=-2. \\
\end{aligned}
\end{equation}

\subsection{Transverse variations}

The concentration fields in the channel undergo a transition from an initially inhomogeneous distribution to a homogenized distribution over long timescales. To study the dynamics of this transition, which cannot be captured by the homogenization calculation, we begin by performing numerical simulations of the governing equation \eqref{eq:Advection-Nernst-Planck equation no potential}, which reveal that the system undergoes complex behavior as it approaches the homogenized state.

\begin{figure}
  \centering
      \includegraphics[width=0.46\linewidth]{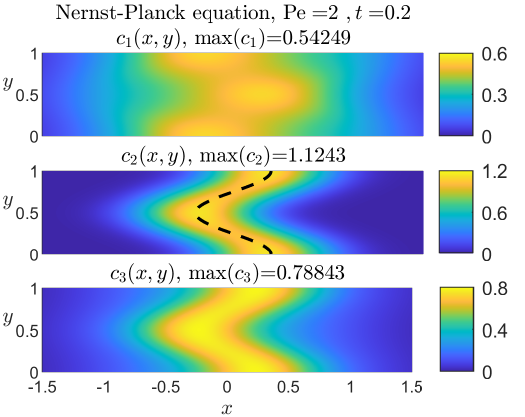}
    \includegraphics[width=0.46\linewidth]{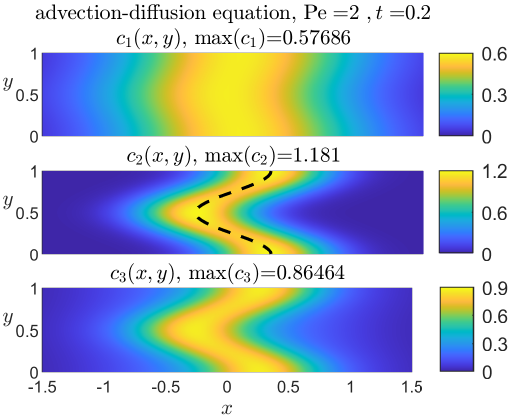}
  \hfill
  \caption[]
  {Numerical solution to equation  \eqref{eq:Advection-Nernst-Planck equation no potential} (left panel) and \eqref{eq:Advection Diffusion} (right panel) at $t=0.2$ with $\mathrm{Pe}=2$. The initial condition, diffusivities and valences are provided in equation \eqref{eq:numerical simulation initial conditoin}. The black dashed lines depict the shape of the shear flow profile. }
  \label{fig:comparision AD ANP}
\end{figure}

The left panel of figure \ref{fig:comparision AD ANP} shows the solution of equation \eqref{eq:Advection-Nernst-Planck equation no potential} at an early stage, $t=0.2$. For comparison, the right panel of Figure \ref{fig:comparision AD ANP} presents the result when the electric potential is negligible, i.e., the solution of the advection-diffusion equation \eqref{eq:Advection Diffusion}. When the simulation time is small compared to the diffusion timescale, the shear flow advection dominates, and one would expect the concentration field to follow the shear flow profile, as shown in the right column of figure \ref{fig:comparision AD ANP}. In the left column, the concentration fields of the second and third ion species also follow the shear flow profile. However, for the first ion species, the concentration field does not follow the expected behavior and instead bends in the opposite direction to the shear flow, as shown in the middle-left plot of Figure \ref{fig:comparision AD ANP}. This behavior is the result of the electric interaction between ions. Both the first and second ion species are cations, so the repulsive electromagnetic force pushes ions away from each other. If one of them follows the shear flow profile, the other one will bend in the opposite direction. As a result, the second ion species visually migrates upstream.

\begin{figure}
  \centering
      \includegraphics[width=0.46\linewidth]{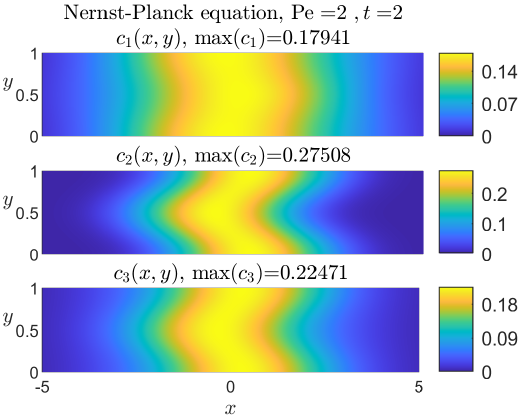}
    \includegraphics[width=0.46\linewidth]{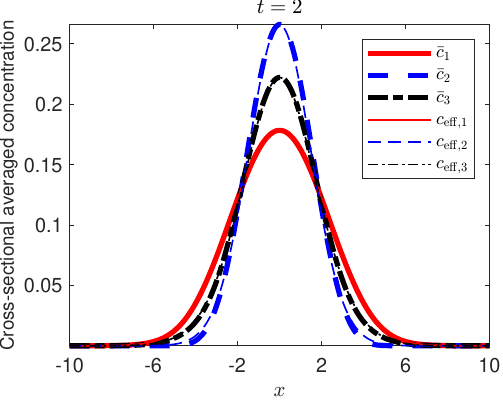}
  \hfill
  \caption[]
  {The left panel shows the numerical solution to equation  \eqref{eq:Advection-Nernst-Planck equation no potential} at $t=2$ with $\mathrm{Pe}=2$. The initial condition, diffusivities and valences are provided in equation \eqref{eq:numerical simulation initial conditoin}. In the right panel, the thicker curves represent the cross-sectional average of concentration fields presented in the left panel, $ \bar{c} (x,t) = \frac{1}{\left| \Omega\right|}  \int\limits_{\Omega}^{}c (x,\mathbf{y},t) \mathrm{d}\mathbf{y}$. The thinner curve represents the solution of effective equation \eqref{eq:homogenization steady effective equation cos}.  The complete dynamics of the simulation in left panel can be observed in Supplementary Movie 1. }
  \label{fig:comparision ANP self similarity solution}
\end{figure}

It is interesting to see how concentration distribution changes at larger time scales where diffusion has a greater influence. The left panel of figure \ref{fig:comparision ANP self similarity solution} presents the numerical solution for equation \eqref{eq:Advection-Nernst-Planck equation no potential} at a larger time $t=2$. As expected, all concentration profiles are more blurred due to diffusion. The concentration profiles of the first and second ions still bend in opposite directions. On the right panel, we compare the cross-sectional averaged concentration field with the solution to the effective equation \eqref{eq:homogenization steady effective equation cos} that was derived using the homogenization method. The curves perfectly overlap, demonstrating the validity of the homogenization calculation. It's worth noting that due to the assumption of the asymptotic analysis, the effective equation is valid for small $\epsilon$ or large $t$, where $\epsilon=\frac{L_{y}}{L_{x}}$ and $L_x$ and $L_y$ are characteristic lengths of the initial condition and the channel width, respectively. In this numerical test case, we have $L_{x}=\frac{1}{4}$, $L_{y}=1$ and $\epsilon=4$. The diffusion time scale is $ \max (\kappa_{1}^{-1},\kappa_{2}^{-1},\kappa_{3}^{-1})=10$, indicating that the parameter regime for the effective equation to reach a good approximation is larger than previously thought.

\begin{figure}
  \centering
 \includegraphics[width=0.46\linewidth]{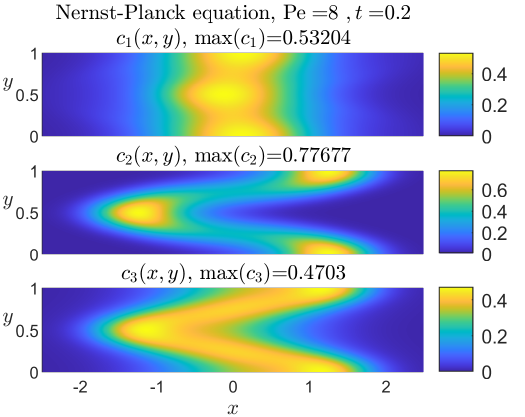}
  \includegraphics[width=0.46\linewidth]{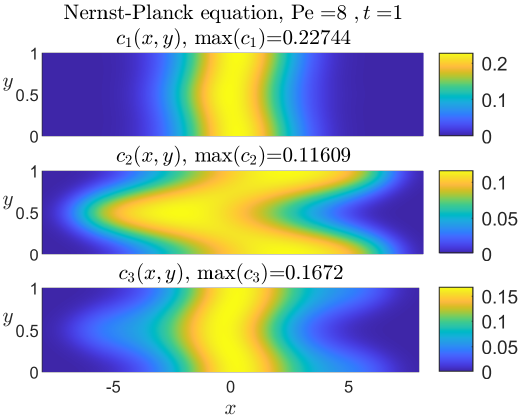}
  \hfill
  \caption[]
  {The numerical solution to equation  \eqref{eq:Advection-Nernst-Planck equation no potential} with $\mathrm{Pe}=8$ at $t=0.2$ (left panel), $t=1$ (right panel).  The initial condition, diffusivities and valences are provided in equation \eqref{eq:numerical simulation initial conditoin}.  The complete dynamics of the simulation can be observed in Supplementary Movie 2. }
  \label{fig:AdvectionNernstPlanckPe8}
\end{figure}

The variations of the concentration field across the channel are different for a stronger flow. Figure \ref{fig:AdvectionNernstPlanckPe8} presents the numerical solutions to the advection-Nernst-Planck equation \eqref{eq:Advection-Nernst-Planck equation no potential} for a stronger flow with $\mathrm{Pe}=8$ at $t=0.2$ (left panel) and $t=1$ (right panel). We have several observations. Firstly, the effect of flow becomes more prominent over the ion-electric interaction, resulting in all concentration profiles bending in the direction of the shear flow. This is in contrast to the case of weak flow, where the concentration profiles of the first and second ion species bend in opposite directions due to the ion-electric interaction. Secondly, at the early stages, there is a clearer separation between the first and second ion species, with the majority of the first ion species remaining near their original positions, while the second ion species are pushed away by the electromagnetic force. However, due to diffusion, the solutions homogenize and the separation becomes weaker as time increases. This homogenization is evident in the right panel of figure \ref{fig:AdvectionNernstPlanckPe8}, where the separation is no longer visible. The third observation is that the different ion species have different spreading rates in the longitudinal direction. The second ion species, with the smallest diffusivity, spreads the most. These unexpected results highlight the importance of studying effective diffusivity in understanding the transport of ions in microchannels under flow conditions.

\subsubsection{Dependence of the effective diffusivity on P\'{e}clet numbers}
\begin{figure}
  \centering
    \subfigure[]{
      \includegraphics[width=0.46\linewidth]{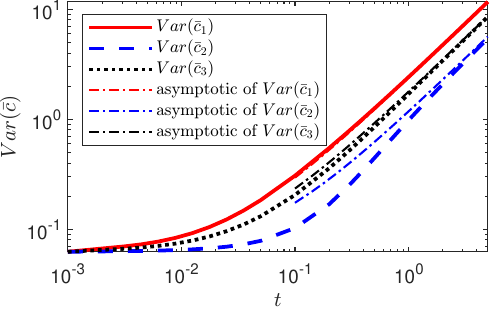}}
      \subfigure[]{
  \includegraphics[width=0.46\linewidth]{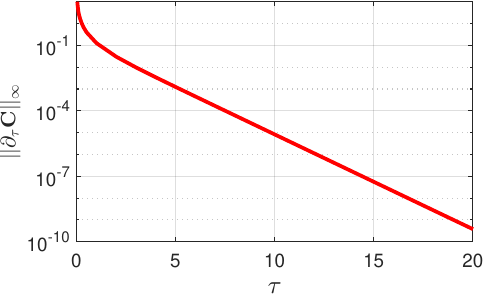}}
  
  \hfill
  \caption[]
  {(a) Comparison between the variance of the numerical solutions for \eqref{eq:Advection-Nernst-Planck equation no potential} (thicker curves) and their theoretical asymptoticsin (thinner curves) log-log scale. The numerical solutions at $t=2$ are plotted in figure \ref{fig:comparision ANP self similarity solution}. The asymptotics of the variance is provided in equation \eqref{eq:variance asymptotics}. In this case, $\mathrm{Var} (\bar{c}_{i})\approx 2 t\kappa_{\mathrm{eff},i} + \sigma^{2}$,  where $\sigma= \frac{1}{4}$ and $\kappa_{\mathrm{eff},i}$ is defined in equation \eqref{eq:diffusivity by similarity solution}.  (b) The infinity norm of $\partial_{\tau}\mathbf{C}$ as a function of $\tau$. $\mathbf{C}$ solves equation  \eqref{eq:homogenization steady effective equation similarity variables0}.  }
  \label{fig:VarianceAndEffectiveDiffusivity}
\end{figure}

In this subsection, we will further explore the dependence of the variance of the cross-sectional-averaged concentration and the effective diffusivity on the  P\'{e}clet numbers. Panel (a) of figure \ref{fig:VarianceAndEffectiveDiffusivity} compares the variance of the longitudinal distribution of the numerical solution with the theoretical variance asymptotics for  $\mathrm{Pe}=2$ and the parameters provided in equation \eqref{eq:numerical simulation initial conditoin}. At a larger time scale, the variance grows linearly, and converges to the asymptotics expansions, demonstrating the validity of the asymptotic analysis.

To calculate the effective diffusivities defined in equation \eqref{eq:effectiveDiffusivityDefinition}, we approximate them using the derivative of $Var (\bar{c}_{i})$ at $t=5$, resulting in $\kappa_{\mathrm{eff},1}=1.1614$, $\kappa_{\mathrm{eff},2}=0.55829$, $\kappa_{\mathrm{eff},3}=0.85984$. On the other hand, equation \eqref{eq:diffusivity by similarity solution} allows us to calculate the effective diffusivity via the self-similarity solution. Notice that the self-similarity solution is the steady solution of equation \eqref{eq:homogenization steady effective equation similarity variables0}. Therefore, we can obtain the self-similarity solution by solving the initial value problem \eqref{eq:homogenization steady effective equation similarity variables0} until the solution reaches a steady state. Panel (b) of figure \ref{fig:VarianceAndEffectiveDiffusivity} plots the infinity norm of $\partial_{\tau}\mathbf{C}$ as a function of $\tau$, which verifies that the solution of the initial value problem converges to the self-similarity solution. Integrating the self-similarity solution yields $\kappa_{\mathrm{eff},1}=1.1618$, $\kappa_{\mathrm{eff},2}=0.558978$, $\kappa_{\mathrm{eff},3}=0.860387$.  The effective diffusivities calculated by two different methods are consistent, demonstrating that self-similarity can accurately characterize the system's dynamics at long times. As an additional verification, we also solve the equivalent equation \eqref{eq:homogenization steady effective equation similarity variables order 1} for the self-similarity solution using the \texttt{NDSolve} in Mathematica, and the results are consistent up to 6 significant digits.

\begin{figure}
  \centering
 \subfigure[]{
 \includegraphics[width=0.46\linewidth]{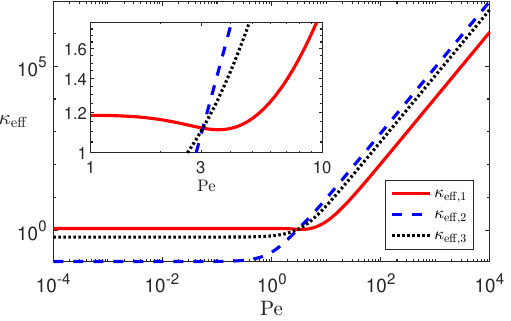}}
\subfigure[]{
\includegraphics[width=0.46\linewidth]{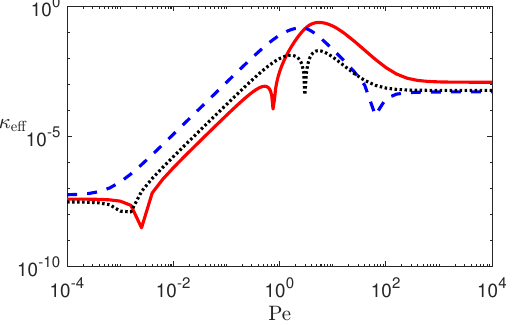}}
 \hfill
  \caption[]
  {(a) The effective diffusivities as a function of the P\'{e}clet number $\mathrm{Pe}$. The inset figure show the function for the P\'{e}clet number from 1 to 10. The three curves intersect at a P\'{e}clet number around 3. (b)The relative difference between the effective diffusivity and its approximation provided in equation \eqref{eq:approximation effective diffusivity ex1}.  The solid red curve represents the first ion species, the dashed blue curve represents the second ion species, and the dotted black curve represents the third ion species. }
  \label{fig:EffectiveDiffusivityPe1}
\end{figure}

Figure \ref{fig:EffectiveDiffusivityPe1} panel (a) shows the effective diffusivities as a function of the P\'{e}clet number $\mathrm{Pe}$ for the parameters provided in equation \eqref{eq:numerical simulation initial conditoin}. We make three observations. First, in classical Taylor dispersion given by equation \eqref{eq:Advection Diffusion effective equation}, the effective diffusivity monotonically increases with the P\'{e}clet number. In contrast, when considering the diffusion-induced electric potential, the effective diffusivities of some ion species may exhibit non-monotonic behavior with respect to the P\'{e}clet number, as shown in the inset of Figure \ref{fig:EffectiveDiffusivityPe1}. Second, the species exhibiting the largest effective diffusivity can vary depending on the P\'{e}clet number. For instance, at small P\'{e}clet numbers, the first ion species demonstrates the highest effective diffusivity. However, when the P\'{e}clet number is approximately three, all three ion species exhibit the same effective diffusivity. Conversely, at large P\'{e}clet numbers, the second ion species displays the greatest effective diffusivity. These findings align with the observations depicted in figures \ref{fig:comparision ANP self similarity solution} and \ref{fig:AdvectionNernstPlanckPe8}. Third, for large P\'{e}clet numbers, the effective diffusivity scales as $\mathrm{Pe}^{2}$, which is the same as the classical Taylor dispersion.

Using the exact solution \eqref{eq:exact solution non zero Pe ex1} and \eqref{eq:exact solution ex1}, the approximation of effective diffusivities \eqref{eq:approximation effective diffusivity} becomes 
\begin{equation}\label{eq:approximation effective diffusivity ex1} 
\begin{aligned}
  &\kappa_{\mathrm{eff},1}=1.1774  \kappa_{1}+ 0.84036 \frac{ \mathrm{Pe}^{2}}{8\pi^{2} \kappa_{1}},\quad \kappa_{\mathrm{eff},2}=1.17514  \kappa_{2}+ 0.71643 \frac{ \mathrm{Pe}^{2}}{8\pi^{2} \kappa_{2}},\\
  &\kappa_{\mathrm{eff},3}=0.64746  \kappa_{3}+ 4.00232 \frac{ \mathrm{Pe}^{2}}{8\pi^{2} \kappa_{3}}.
\end{aligned}
\end{equation}
Panel (b) of figure \ref{fig:EffectiveDiffusivityPe1} displays the relative difference between the effective diffusivity and its approximation, revealing that the approximation performs well for both small and large Peclet numbers.  In a microfluidic experiment, a typical flow speed can be 0.2 cm/s, and the channel width is 0.05 cm, the diffusivity of the solute is around $10^{-5}$ cm$^{2}$/s,  the resulting in a P\'eclet number of 1000. The aforementioned approximation performs well in this parameter regime.

We are interested in comparing these results with two other ``naive'' approaches. In the first approach, we neglect the diffusion-induced electric potential and assume that all ions are passively advected by the shear flow. As a result, the governing equation simplifies to the advection-diffusion equation. In this case, the effective diffusivities for each ion species can be calculated as follows:
\begin{equation}\label{eq:approximation effective diffusivity ex2} 
\begin{aligned}
&\kappa_{\mathrm{eff},i}=\kappa_{i}+ \frac{ \mathrm{Pe}^{2}}{8\pi^{2} \kappa_{i}}, \quad i=1,2,3.
\end{aligned}
\end{equation}

In the second approach, we assume that the solution is a mixture of two binary electrolytes. We further assume that there is no interaction between these two electrolytes, and they are passively advected by the shear flow. The first binary electrolyte consists of the first and third ion species, while the second binary electrolyte consists of the second and third ion species. 
By employing the formula for binary electrolytes as presented in Equation \eqref{eq:homogenization steady effective equation two ions}, we can determine the diffusivity of the first and second binary electrolytes to be 1 and $\frac{1}{7}$, respectively. In this case, the effective diffusivities for each ion species can be calculated as follows:
\begin{equation}\label{eq:approximation effective diffusivity ex3} 
\begin{aligned}
  &\kappa_{\mathrm{eff},1}=\kappa_{1}+ \frac{ \mathrm{Pe}^{2}}{8\pi^{2} \kappa_{1}}, \quad \kappa_{\mathrm{eff},2}=\frac{10}{7}\kappa_{2}+ \frac{7}{10}\frac{ \mathrm{Pe}^{2}}{8\pi^{2} \kappa_{2}}\approx 1.4286\kappa_{2}+ 0.7\frac{ \mathrm{Pe}^{2}}{8\pi^{2} \kappa_{2}},\\
&\kappa_{\mathrm{eff},3}=\frac{4}{7}\kappa_{3}+ 4\frac{ \mathrm{Pe}^{2}}{8\pi^{2} \kappa_{3}}\approx 0.5714\kappa_{3}+ 4\frac{ \mathrm{Pe}^{2}}{8\pi^{2} \kappa_{3}}.
\end{aligned}
\end{equation}

In this example, equation \eqref{eq:approximation effective diffusivity ex2} fails to provide a good estimation for the effective diffusivity of the three ion species. As we expected, since equation \eqref{eq:approximation effective diffusivity ex3} takes into account the ion-electric interaction in each binary electrolyte,  equation \eqref{eq:approximation effective diffusivity ex3} is closer to equation \eqref{eq:approximation effective diffusivity ex1} than equation \eqref{eq:approximation effective diffusivity ex2}. Both equation \eqref{eq:approximation effective diffusivity ex2} and equation \eqref{eq:approximation effective diffusivity ex3} underestimate the effective diffusivity of the first ion species for small Peclet numbers and overestimate it for large Peclet numbers compared to equation \eqref{eq:approximation effective diffusivity ex1}. Interestingly, for the effective diffusivity of the second and third ion species, equation \eqref{eq:approximation effective diffusivity ex3} differs from equation \eqref{eq:approximation effective diffusivity ex1} for small P\'eclet numbers, while it is very close to equation \eqref{eq:approximation effective diffusivity ex2} for large P\'eclet numbers. Therefore, treating the electrolytes in mixture independently may describe the effective diffusivity for some ion species with a reasonable error, but it may not correctly describe the effective diffusivity for all ion species.

Finally, it is worth noting that the effective diffusivity resulting from the initial conditions of the same type is the same. However, when the initial conditions belong to different types, even if the physical parameters and mass ratio are the same, the resulting effective diffusivity can be different. For instance, when the initial conditions are given by $c_{I,1}=c_{I,1}= \frac{1+ \mathrm{erf}(x/2)}{2}$, with $\mathrm{Pe}=2$ and the same diffusivities and valences provided in equation \eqref{eq:numerical simulation initial conditoin}, the effective diffusivities are $\kappa_{\mathrm{eff},1}=1.1554$, $\kappa_{\mathrm{eff},2}=0.55242$, and $\kappa_{\mathrm{eff},3}=0.85392$. This result differs from the case when the initial condition is the Gaussian distribution function. While the relative difference in effective diffusivity between the initial conditions considered here is small, it is worth noting that this may not always be the case. In other parameter regimes, the difference in effective diffusivity between different initial conditions could be more significant.

\subsection{Ion separation}
\label{sec:ion separation}
After the solute has been homogenized across the channel, the concentration distribution is described by a self-similar solution of the effective equation \eqref{eq:homogenization steady effective equation similarity variables}.  In some parameter regimes, this self-similarity solution exhibiting properties that differ from the case without the diffusion-induced electric potential. Here, we examine the shape of this solution and explore these unique properties in more details.

\begin{figure}
  \centering
  \subfigure[]{
    \includegraphics[width=0.46\linewidth]{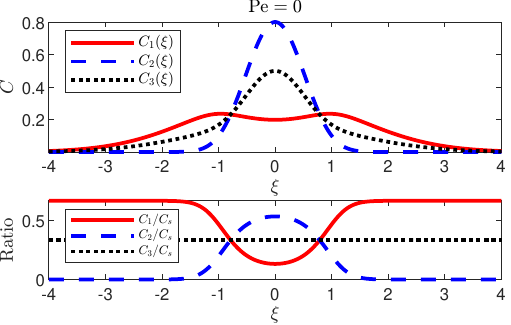}
  }
\subfigure[]{
    \includegraphics[width=0.46\linewidth]{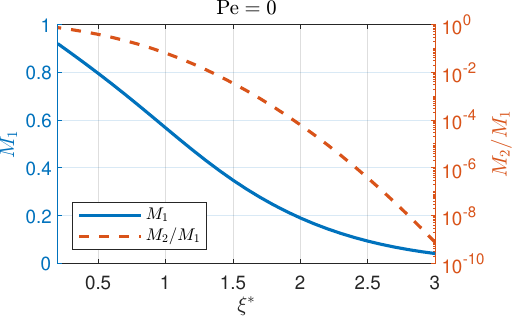}
  }  
  \caption[]
  {(a) The upper panel shows numerically solved self-similarity solution of the effective equation \eqref{eq:homogenization steady effective equation similarity variables}. The initial condition, diffusivities and valences are provided in equation \eqref{eq:numerical simulation initial conditoin}.       The lower panel shows the ratio of each component, $C_{i}/\sum\limits_{i=1}^{3}C_{i}$. (b) The blue solid curve represents $M_{1} (\xi^{*})$ and is associated with the left y axis. The read dashed curve represents $M_{2}(\xi^{*})/M_{1}(\xi^{*})$ and is associated with the right y axis.  }
  \label{fig:SelfSimilaritySolutionPe0}
\end{figure}

The upper panel of figure \ref{fig:SelfSimilaritySolutionPe0} (a) displays the self-similarity solution $C_{i}(\xi)$ for $\mathrm{Pe}=0$, where $\xi=\frac{x}{\sqrt{t}}$. Interestingly, $C_{1} (\xi)$ exhibits a highly non-Gaussian shape and is not even unimodal.   It's important to mention that deviations from regular Gaussian profiles are due to the ion-electric interaction and can be observed without relying on the shear flow. In the lower panel of figure \ref{fig:SelfSimilaritySolutionPe0} (a), we plot the ratio of each component ${C_{i}}/{\sum\limits_{i=1}^3 C_{i}}$ as a function of $\xi$. The ratio of the third ion species is almost constant, while the ratios of the first and second ion species vary significantly. At small values of $\xi$, there are more second ion species than first ion species, while at large $\xi$, there are virtually no second ion species in the relative sense. These results imply a spontaneous separation of ions.

By keeping the solution to the region $|\xi| > \xi^{*}$ for some threshold $\xi^{*}$, which is practical for experimental implementation, we can obtain a solution that consists predominantly of the first and third ions, which implies that we can separate one binary electrolyte from the mixture of three ion species. To qualitatively investigate the ion separation, we plot the mass of each component $M_{i}(\xi^{*}) = \int_{-\infty}^{-\xi^{*}} C_{i}(\xi) d\xi + \int_{\xi^{*}}^{\infty} C_{i}(\xi) d\xi$ and their ratio $\frac{M_{2}}{M_{1}}$ as functions of $\xi^{*}$ in figure \ref{fig:SelfSimilaritySolutionPe0} panel (b). For example, by allowing a tolerance ratio of $M_{2}/M_{1}=0.1$, we can choose $\xi^{*} \approx 0.9003$, which keeps the mass $M_1=0.6145$. Note that this method can separate approximately 61\% of the binary electrolytes consisting of the first and third ion species from the mixture of three ion species, given that the total mass of the first ion species is 1. If the  tolerance ratio decreases to $M_{2}/M_{1}=0.01$, we can choose $\xi^{*} \approx 1.3409$, which still retains 41\% of the first ion species.

\begin{figure}
  \centering
  \subfigure[]{
    \includegraphics[width=0.46\linewidth]{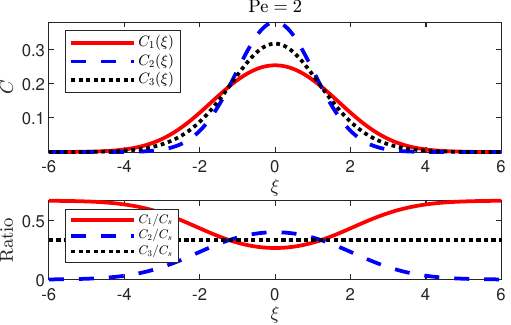}
  }
  \subfigure[]{
    \includegraphics[width=0.46\linewidth]{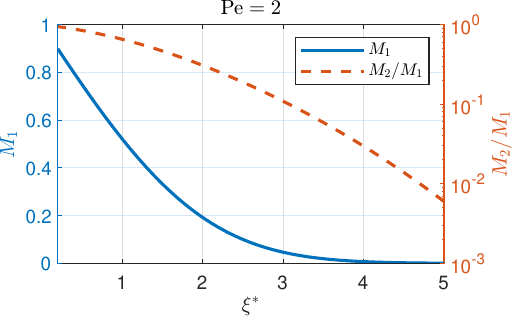}
  }
  \hfill
  \caption[]
  { (a)The upper panel shows  self-similarity solution of the effective equation \eqref{eq:homogenization steady effective equation similarity variables} with  $\mathrm{Pe}=2$. The initial condition, diffusivities and valences are provided in equation \eqref{eq:numerical simulation initial conditoin}.   The lower panel shows the ratio of each component, $C_{i}/\sum\limits_{i=1}^{3}C_{i}$. (b) The blue solid curve represents $M_{1} (\xi^{*})$ and associate with the left y axis. The read dashed curve represents $M_{2}(\xi^{*})/M_{1}(\xi^{*})$ and associate with the right y axis.  }
  \label{fig:SelfSimilaritySolutionPe2}
\end{figure}

In the presence of a shear flow, the ion separation may be weakened. Figure \ref{fig:SelfSimilaritySolutionPe2} summarizes the results for the case where the P\'{e}clet number is $\mathrm{Pe}=2$. In this case, the concentration profiles $C_i$ become unimodal functions that are close to Gaussian distributions. Although there are still very few second ion species present in the solution for large $\xi^{*}$, the amount of first ion species that can be retained through separation is much smaller compared to the case without flow. To illustrate this, we consider a tolerance ratio of $M_{2}/M_{1}=0.1$. The optimal value of $\xi^{*}$ that achieves this ratio is found to be $\xi^{*}=3.0741$, and the mass of the separated first ion species is $M_1=0.04121$. If we reduce the tolerance ratio to $M_{2}/M_{1}=0.01$, the optimal value of $\xi^{*}$ is approximately $\xi^{*} \approx 4.7011$, but the mass of the separated first ion species is much smaller, with $M_1=0.001477$. These results suggest that the separation of different ion species is weaker in the presence of a shear flow, indicating that the flow strength plays an important role in the separation process.

This example suggests that the presence of a shear flow can weaken the separation of different ion species, highlighting the importance of flow strength in the separation process. On the other hand, this can be rationalized by observing figure \ref{fig:EffectiveDiffusivityPe1}, where the effective diffusivities of the three ion species converge to the same value around $\mathrm{Pe} = 3$. As $\mathrm{Pe}$ approaches 3, the dispersion rates of the three ion species become closer, resulting in weaker separation. However, when the P\'{e}clet number increases beyond 3, the difference in effective diffusivities increases, which results  stronger separation.  The right column of figure \ref{fig:AdvectionNernstPlanckPe8} illustrates this phenomenon, where the second ion species exhibits the highest dispersion. Consequently, at larger values of $x$, the solution predominantly consists of the second and third ions. This effect becomes even more pronounced at higher P\'{e}clet numbers.

\begin{figure}
  \centering
  \subfigure[]{
    \includegraphics[width=0.46\linewidth]{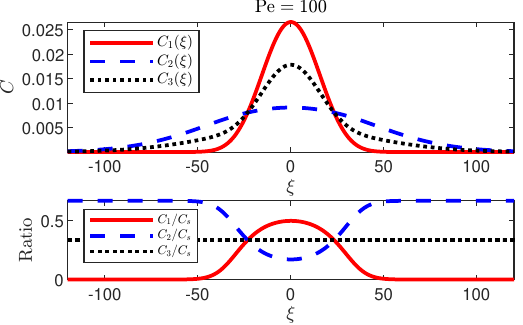}
  }
  \subfigure[]{
    \includegraphics[width=0.46\linewidth]{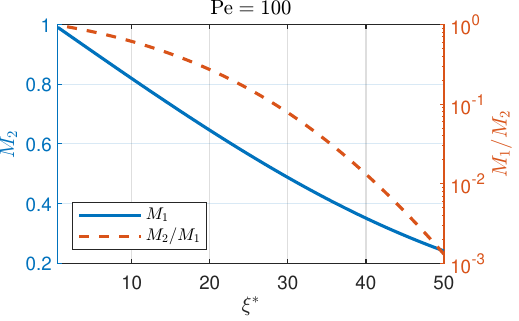}
  }
  \caption[]
  {  (a)The upper panel shows  self-similarity solution of the effective equation \eqref{eq:homogenization steady effective equation similarity variables}  with $\mathrm{Pe}=100$.  The initial condition, diffusivities and valences are provided in equation \eqref{eq:numerical simulation initial conditoin}. The lower panel shows the ratio of each component, $C_{i}/\sum\limits_{i=1}^{3}C_{i}$.  (b) The blue solid curve represents $M_{2} (\xi^{*})$ and associate with the left y axis. The read dashed curve represents $M_{1}(\xi^{*})/M_{2}(\xi^{*})$ and associate with the right y axis.   }
  \label{fig:SelfSimilaritySolutionPe100}
\end{figure}

Panel (a) of figure \ref{fig:SelfSimilaritySolutionPe100} illustrates the self-similarity solution for $\mathrm{Pe} = 100$, using identical diffusivities and valences. Surprisingly, the result is contrary to that observed when $\mathrm{Pe} = 0$. Specifically, at small values of $\xi$, the first ion species outnumber the second ion species, whereas at large $\xi$, the relative abundance of first ion species becomes negligible. In Figure \ref{fig:SelfSimilaritySolutionPe100} panel (b) displays  $M_{2}(\xi^{})$ and $\frac{M_{1}}{M_{2}}$ as functions of $\xi^{*}$. In this scenario, it is possible to obtain a solution that mainly consists of the second and third ion species by retaining the solution at $\xi > \xi^{*}$, while still preserving a reasonable amount of the second ion species. For instance, if we allow a tolerance ratio of $M_{2}/M_{1} = 0.1$, we can select $\xi^{*} \approx -28.3$, resulting in a mass of $M_2 = 0.51308$. Notably, this approach allows the separation of approximately 61\% of the binary electrolytes comprising the first and third ion species from the mixture of three ion species, considering that the total mass of the first ion species is 1. If we reduce the tolerance ratio to $M_{2}/M_{1} = 0.01$, we can choose $\xi^{*} \approx -41.33$, which still retains 33.5\% of the first ion species.

\begin{figure}
  \centering
    \includegraphics[width=0.46\linewidth]{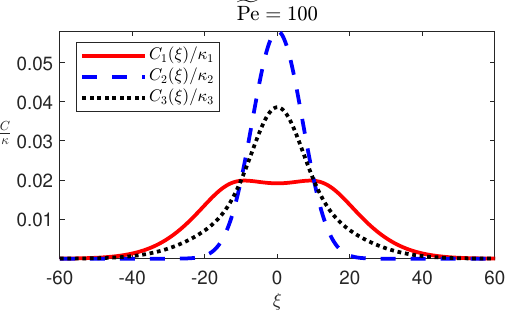}
  \hfill
  \caption[]
  {The normalized self similarity solution for effective equation \eqref{eq:homogenization steady effective equation similarity variables}  with $\widetilde{\mathrm{Pe}}=100$ and the  diffusivities  $\tilde{\kappa}_{1}=1$, $\tilde{\kappa}_{2}=10$, $\tilde{\kappa}_{3}=1$,  valances $\tilde{z}_{1}=1$, $\tilde{z}_{2}=0.1$, $\tilde{z}_{3}=-2$ (or equivalent $\tilde{z}_{1}=10$, $\tilde{z}_{2}=1$, $\tilde{z}_{3}=-20$), and the initial condition $\tilde{c}_{I,1}=\frac{1}{\sigma\sqrt{2\pi}} e^{-\frac{1}{2} \left( \frac{x}{\sigma} \right)^{2}}$, $\tilde{c}_{I,2}=\frac{10}{\sigma\sqrt{2\pi}} e^{-\frac{1}{2} \left( \frac{x}{\sigma} \right)^{2}}$, where $\sigma=\frac{1}{4}$.  }
  \label{fig:SelfSimilaritySolutionPe100case2}
\end{figure}

Finally, the reciprocal property discussed in Section \ref{sec:Comparison to the Taylor dispersion} suggests that the observed phenomena in this specific system could also manifest in different systems. To illustrate this, we present an example that highlights the reciprocal property and demonstrates how ion separation, observed in the aforementioned system without fluid flow, can occur in a different system characterized by strong fluid flow. By employing the change of variable described in Section \ref{sec:Comparison to the Taylor dispersion}, we establish an equivalence between the system discussed previously and the system with  $\tilde{\kappa}_{1}=1$, $\tilde{\kappa}_{2}=10$, $\tilde{\kappa}_{3}=1$, $\tilde{z}_{1}=1$, $\tilde{z}_{2}=0.1$, $\tilde{z}_{3}=-2$, and initial conditions of $\tilde{c}_{I,1}=\frac{1}{\sigma\sqrt{2\pi}} e^{-\frac{1}{2} \left( \frac{x}{\sigma} \right)^{2}}$ and $\tilde{c}_{I,2}=\frac{10}{\sigma\sqrt{2\pi}} e^{-\frac{1}{2} \left( \frac{x}{\sigma} \right)^{2}}$, where $\sigma=\frac{1}{4}$. We proceed to numerically solve the self-similarity solution of the transformed system with $\widetilde{\mathrm{Pe}}=100$ and present it in Figure \ref{fig:SelfSimilaritySolutionPe100case2}. According to the reciprocal property, the normalized solution should be equivalent to the solution of the original system with $\mathrm{Pe}=\frac{1}{100}$. In fact, the solution of the transformed system closely resembles the solution depicted in Figure \ref{fig:SelfSimilaritySolutionPe0}, demonstrating that the highly non-Gaussian shape and lack of unimodality can exist in the presence of strong flow.

\subsection{limiting of concentration}  
The concentration-dependent diffusion-induced electric potential gives rise to variations in the effective diffusivities of the ion species. Table \ref{tab:Effective diffusivity for different ratio} presents the effective diffusivities for different mass ratios of the second and first ion species $\int\limits_{-\infty}^{\infty} c_{2}\mathrm{d}x/ \int\limits_{-\infty}^{\infty} c_{1}\mathrm{d} x$ with fixed diffusivities and valences: $\kappa_{1}=1$, $\kappa_{2}=0.1$, $\kappa_{3}=1$, $z_{1}=1$, $z_{2}=1$, and $z_{3}=-2$, and two different P\'{e}clet numbers, $\mathrm{Pe}=0$ and 2. . The table shows the triplet of effective diffusivities $(\kappa_{\mathrm{eff},1},\kappa_{\mathrm{eff},2},\kappa_{\mathrm{eff},3})$ vary widely,  as the mass ratio increases from 0.01 to 100.

 \begin{table}
  \begin{center}
\def~{\hphantom{0}}
  \begin{tabular}{l|l|ccccccc}
   & Ratio    &0.01  & 0.1 & 0.5   &   1 & 2& 10 &100 \\
    \hline
$\mathrm{Pe}=0$&$\kappa_{\mathrm{eff},1}$&1.00297 &1.02772 &1.10914 &1.1774 &1.26774 &1.5464 &2.0527\\
&$\kappa_{\mathrm{eff},2}$&0.100359 &0.103251 & 0.111673 &0.117514&0.12373&0.13505& 0.14135\\
&$\kappa_{\mathrm{eff},3}$&0.994038 &0.943682 & 0.776652 &0.647457 &0.50507&0.26336 &0.16028\\
     \hline
$\mathrm{Pe}=2$&$\kappa_{\mathrm{eff},1}$&1.05214 &1.06501 & 1.11424 &1.16179 &1.22965 &1.44014&1.74591\\
&$\kappa_{\mathrm{eff},2}$&0.605669 &0.598014 &0.57506 &0.558975&0.542145&0.513617 &0.500019  \\
    &$\kappa_{\mathrm{eff},3}$&1.04772 &1.02256 & 0.93452 &0.860384 &0.771314&0.597846 &0.512355\\
        \hline
  \end{tabular}
  \caption{Effective diffusivity for different mass ratios of the second and first ion species $\int\limits_{-\infty}^{\infty} c_{2}\mathrm{d}x/ \int\limits_{-\infty}^{\infty} c_{1}\mathrm{d} x$.  The parameters are  $\kappa_{1}=1, \kappa_{2}=0.1, \kappa_{3}=1, z_{1}=1, z_{2}=1, z_{3}=-2$. }
  \label{tab:Effective diffusivity for different ratio}
  \end{center}
\end{table}

As the mass ratio of the second and first ion species decreases to zero, the solution becomes dominated by the first and third ion species, and the effective diffusivity can be calculated using the formula for binary electrolytes in equation
 \eqref{eq:homogenization steady effective equation two ions},
\begin{equation}
\begin{aligned}
 &\kappa_{\mathrm{eff},1}=\kappa_{\mathrm{eff},3}=\frac{\kappa _1 \kappa _3 \left(z_1-z_3\right)}{\kappa _1 z_1-\kappa _3 z_3}+\frac{\mathrm{Pe}^2}{8\pi^{2}}\frac{\kappa _1 z_1-\kappa _3 z_3}{\kappa _1 \kappa _3 \left(z_1-z_3\right)}=1+\frac{1}{2 \pi ^2}\approx 1.05066. \\
\end{aligned}
\end{equation}
As shown previously in figure \ref{fig:SelfSimilaritySolutionPe0} and \ref{fig:SelfSimilaritySolutionPe2}, for these physical parameters, $c_{1}$ is much larger than $c_{2}$ for large $x$. Therefore, the mass ratio of the second and first ion species decreasing to zero implies that $c_{2}/c_{1}\rightarrow 0$ for all $x$. As a result, the diffusion tensor  $\mathbf{D}+\mathrm{Pe}^{2}\mathbf{D}^{-1}$ provided in \eqref{eq:diffussion tensor three ions} becomes
\begin{equation}
  \begin{aligned}
    \begin{bmatrix}
  \frac{\kappa _1 \kappa _3 \left(z_1-z_3\right)}{\kappa _1 z_1-\kappa _3 z_3}+\frac{\mathrm{Pe}^2}{8\pi^{2}}\frac{\kappa _1 z_1-\kappa _3 z_3}{\kappa _1 \kappa _3 \left(z_1-z_3\right)}& \frac{\mathrm{Pe}^2}{8\pi^{2}} \frac{\left(\kappa _2-\kappa _3\right)  z_2}{\kappa _2 \kappa _3 \left(z_1-z_3\right)}-\frac{\kappa _1 \left(\kappa _2-\kappa _3\right) z_2}{\kappa _1 z_1-\kappa _3 z_3} \\
 0 & \kappa _2+\frac{\mathrm{Pe}^2}{8 \pi^{2}\kappa _2} \\
    \end{bmatrix}.
\end{aligned}
\end{equation}
Therefore, the formula of $\kappa_{\mathrm{eff},2}$ remains  the same as the formula in classical Taylor dispersion \eqref{eq:Advection Diffusion effective equation}
\begin{equation}
\begin{aligned}
\kappa_{\mathrm{eff},2}= &\kappa _2+\frac{\mathrm{Pe}^2}{8 \pi^{2} \kappa _2}=\frac{1}{10}+\frac{5}{\pi ^2}\approx 0.606606, 
\end{aligned}
\end{equation}
which suggests that when the concentration of the second ion species is much smaller than that of the first and third ion species, the second ion species can be considered as passively advected by the shear flow and is decoupled from the first and third ion species.

In the opposite limit, where the mass ratio of the second and first ion species tends to infinity, the effective diffusivities of the second and third ion species converge to 
\begin{equation}
  \begin{aligned}
 &\kappa_{\mathrm{eff},2}=\kappa_{\mathrm{eff},3}=\frac{\kappa _2 \kappa _3 \left(z_2-z_3\right)}{\kappa _2 z_2-\kappa _3 z_3}+\frac{\mathrm{Pe}^2}{8\pi^{2}}\frac{\kappa _2 z_2-\kappa _3 z_3}{\kappa _2 \kappa _3 \left(z_2-z_3\right)}=\frac{1}{7}+\frac{7}{2 \pi ^2}\approx 0.497481, \\
\end{aligned}
\end{equation}
which is consistent with the formula for binary electrolytes in equation  \eqref{eq:homogenization steady effective equation two ions}.
Although we may expect the first ion species still follows the formula of the Taylor dispersion, $\kappa_{\mathrm{eff},1}= \kappa _1+\frac{\mathrm{Pe}^2}{8 \pi^{2} \kappa _1}=1+\frac{1}{2 \pi ^2} \approx 1.05066$,  is inconsistent with the value presented in table \ref{tab:Effective diffusivity for different ratio}. The reason is that the limit  $\int\limits_{-\infty}^{\infty} c_{2}\mathrm{d}x/ \int\limits_{-\infty}^{\infty} c_{1}\mathrm{d} x\rightarrow \infty$ does not necessarily imply  $c_{2}/c_{1}\rightarrow \infty$ uniformly for all $x$ according to the exact solution \eqref{eq:exact solution ex1}, and the conditions for the previous asymptotic analysis are not valid. Due to the nonlinearity of the problem, it is difficult to find a closed-form expression for $\kappa_{\mathrm{eff},1}$ in this limit.

The classical theory of Taylor dispersion has been used to study the effective diffusivity of a solute in a channel, taking into account factors such as molecular diffusivity, channel cross-sectional geometry, and flow rate. Using the measured effective diffusivities, the molecular diffusivity can be easily calculated by determining the geometry factor and flow rate. This technique has been widely used for diffusivity measurement (\cite{bello1994use,taladriz2019precision,leaist2017quinary}). However, as we have demonstrated in this section for a multispecies electrolyte solution, the effective diffusivity is also dependent on the concentration ratio of the components. This finding suggests that the Taylor dispersion method can be used to identify the relative concentrations of the components in a multispecies electrolyte mixture.

\section{Conclusion and discussion}
\label{sec:Conclusion and discussion}
This paper presents a theoretical and numerical study on the interplay between shear flow advection and ion-electric interaction in electrically neutral multispecies electrolyte solutions within a channel domain, without the presence of an external electric field.

The governing equation for this system is the advection-Nernst-Planck equation, denoted as \eqref{eq:Advection-Nernst-Planck equation}. In order to simplify the analysis, we have derived an effective equation using homogenization methods, which captures the behavior of the system at the diffusion time scale or when the length scale of the initial data is much larger than the channel width. For unsteady shear flows, the effective equation is represented by equation \eqref{eq:homogenization unsteady effective equation}, while for steady shear flows, it is given by equation \eqref{eq:homogenization steady effective equation}. Importantly, the effective equation only depends on the longitudinal variable of the channel and time, making it easier to solve compared to the governing equation, while still capturing its essential features. Furthermore, we have included the explicit form of the effective equation for several commonly encountered channel geometries and flow conditions. For instance,  equation \eqref{eq:homogenization steady effective equation parallel plate} is  the effective equation for pressure-driven flow in a parallel plate channel domain. Similarly, equation \eqref{eq:homogenization steady effective equation circular pipe} provides the effective equation for flow in a circular pipe. 

Several conclusions have been drawn from the analysis of the effective equation. Firstly, it has been observed that the solution to the effective equation \eqref{eq:homogenization unsteady effective equation} converges to a self-similarity solution described by equation \eqref{eq:homogenization steady effective equation similarity variables} at long times. By examining the scaling properties of this self-similarity solution, it has been demonstrated that the concentration distribution's variance increases linearly over time. Furthermore, the self-similar solution of the effective equation can be utilized to calculate the effective diffusivity using equation \eqref{eq:diffusivity by similarity solution}. Secondly, it has been shown that the nonlinear effective equation can be approximated by a diffusion equation when the background concentration is nonzero. This approximation provides a formula for measuring the mutual diffusion coefficients. Third, it has been demonstrated that the effective equation exhibits a reciprocal property, meaning that a system with weak flow is equivalent to a system with strong flow and appropriately scaled physical parameters. Furthermore, we obtain the exact self-similarity solution for the effective equation involving three ion species in some cases. When the diffusivities of two ions are equal, the solutions are described in equation \eqref{eq:exact solution same diffusivity} and \eqref{eq:exact solution non zero Pe ex1}.  For a special combination of the valences of the ions, the solution is given in equation \eqref{eq:exact solution ex3}. Last, we provide asymptotic analyses for ions with significant diffusivity discrepancies. When one ion species has an extremely large diffusivity compared to the remaining ions, we offer an asymptotic approximation for the self-similarity solution in equation \eqref{eq:exact solution large diffusivity}. Conversely, when one ion species has an extremely small diffusivity, the effective diffusivity can be approximated using equation \eqref{eq:effective diffusivity small diffusivity}.

In addition to the analytical results, to validate our analytical findings, we have conducted numerical simulations, which reveal several interesting properties arising from the nonlinearity of the advection-Nernst-Planck equation.   Firstly, we observe that ion-electric interaction can dominate over shear flow, resulting in some species moving in the opposite direction of the shear flow. Secondly, different ion species can separate at the early stage or at the diffusion time scale, and the degree of separation can be increased or decreased by the shear flow, depending on the physical parameters.  Thirdly, effective diffusivity can be a non-monotonic function of the P\'{e}clet number, in contrast to Taylor dispersion where the effective diffusivity monotonically increases with the P\'{e}clet number. Fourth, when the initial conditions belong to different types, even if the physical parameters and mass ratio are the same, the resulting effective diffusivity can be different. Fifth, even with the Gaussian initial condition, the longitudinal distribution of the concentration can have a highly non-Gaussian shape and may not be unimodal. Fifth, the relationship between effective diffusivity and concentration offers a method to calculate the ratio of each component's concentrations.

The future study includes several directions. Firstly, while we mainly focused on solutions with three ion species in our numerical simulations, it would be interesting to extend the study to solutions with more components. Secondly, the current study only considers straight channel domains, but the inclusion of curved boundaries would provide insight into many practical applications such as  manufacturing a passive mixer for microchannels (\cite{stone2004engineering,stroock2002chaotic,ajdari2006hydrodynamic,oevreeide2020curved}), modeling the fluid flows over rough surfaces \cite{carney2022heterogeneous}, analyzing solute transport in river  (\cite{fischer1969effect,smith1983longitudinal,yotsukura1976transverse}), modeling blood vessel (\cite{marbach2019active}). Thirdly, while our study considers the scalar passively advected by the fluid flow, future research could explore the full coupling of the ion-electric interaction with the fluid equation such as Nernst-Planck-Euler system (\cite{ignatova2021global}), providing a more comprehensive understanding of the system's behavior.
Lastly, our study primarily explores the system in the absence of an external electric field. However, by introducing an applied external electric field, the electro-osmotic flow becomes significant. This can lead to the emergence of nonlinear macro-transport equations, resulting in non-Gaussian solute profiles (\cite{ghosal2012electromigration, ghosal2010nonlinear}). Additionally, the time-varying external field can generate an asymmetric rectified electric field (\cite{hashemi2018oscillating}), which in turn affects solute transport. Exploring these cases would be of great interest for extending our study.

\section{Acknowledgements}
I would like to acknowledge the inspiration for this study provided by Robert Hunt and Professor Richard M. McLaughlin, who brought the paper  (\cite{gupta2019diffusion}) to my attention. In addition, I thank  anonymous referees, whose comments improved the quality of the manuscript.

\bibliographystyle{jfm}

\begin{thebibliography}{72}
\expandafter\ifx\csname natexlab\endcsname\relax\def\natexlab#1{#1}\fi
\def\au#1{#1} \def\ed#1{#1} \def\yr#1{#1}\def\at#1{#1}\def\jt#1{\textit{#1}}
  \def\bt#1{#1}\def\bvol#1{\textbf{#1}} \def\vol#1{#1} \def\pg#1{#1}
  \def\publ#1{#1}\def\arxiv#1{#1}\def\org#1{#1}\def\st#1{\textit{#1}}

\bibitem[Ajdari {\em et~al.\/}(2006)Ajdari, Bontoux \&
  Stone]{ajdari2006hydrodynamic}
{\sc \au{Ajdari, Armand}, \au{Bontoux, Nathalie} \& \au{Stone, Howard~A}}
  \yr{2006}  \at{Hydrodynamic dispersion in shallow microchannels: the effect
  of cross-sectional shape}.  \jt{Analytical Chemistry}  \bvol{78}~(2),
  \pg{387--392}.

\bibitem[Alessio {\em et~al.\/}(2022)Alessio, Shim, Gupta \&
  Stone]{alessio2022diffusioosmosis}
{\sc \au{Alessio, Benjamin~M}, \au{Shim, Suin}, \au{Gupta, Ankur} \& \au{Stone,
  Howard~A}} \yr{2022}  \at{Diffusioosmosis-driven dispersion of colloids: a
  taylor dispersion analysis with experimental validation}.  \jt{Journal of
  Fluid Mechanics}  \bvol{942},  \pg{A23}.

\bibitem[Aminian {\em et~al.\/}(2016)Aminian, Bernardi, Camassa, Harris \&
  McLaughlin]{aminian2016boundaries}
{\sc \au{Aminian, Manuchehr}, \au{Bernardi, Francesca}, \au{Camassa, Roberto},
  \au{Harris, Daniel~M} \& \au{McLaughlin, Richard~M}} \yr{2016}  \at{How
  boundaries shape chemical delivery in microfluidics}.  \jt{Science}
  \bvol{354}~(6317),  \pg{1252--1256}.

\bibitem[Aminian {\em et~al.\/}(2018)Aminian, Bernardi, Camassa, Harris \&
  McLaughlin]{aminian2018diffusion}
{\sc \au{Aminian, Manuchehr}, \au{Bernardi, Francesca}, \au{Camassa, Roberto},
  \au{Harris, Daniel~M} \& \au{McLaughlin, Richard~M}} \yr{2018}  \at{The
  diffusion of passive tracers in laminar shear flow}.  \jt{JoVE (Journal of
  Visualized Experiments)} ~(135),  \pg{e57205}.

\bibitem[Aminian {\em et~al.\/}(2015)Aminian, Bernardi, Camassa \&
  McLaughlin]{aminian2015squaring}
{\sc \au{Aminian, Manuchehr}, \au{Bernardi, Francesca}, \au{Camassa, Roberto}
  \& \au{McLaughlin, Richard~M}} \yr{2015}  \at{Squaring the circle: Geometric
  skewness and symmetry breaking for passive scalar transport in ducts and
  pipes}.  \jt{Physical review letters}  \bvol{115}~(15),  \pg{154503}.

\bibitem[Aris(1956)]{aris1956dispersion}
{\sc \au{Aris, Rutherford}} \yr{1956}  \at{On the dispersion of a solute in a
  fluid flowing through a tube}.  \jt{Proceedings of the Royal Society of
  London. Series A. Mathematical and Physical Sciences}  \bvol{235}~(1200),
  \pg{67--77}.

\bibitem[Aris(1960)]{aris1960dispersion}
{\sc \au{Aris, R}} \yr{1960}  \at{On the dispersion of a solute in pulsating
  flow through a tube}.  \jt{Proceedings of the Royal Society of London. Series
  A. Mathematical and Physical Sciences}  \bvol{259}~(1298),  \pg{370--376}.

\bibitem[Ault {\em et~al.\/}(2017)Ault, Warren, Shin \&
  Stone]{ault2017diffusiophoresis}
{\sc \au{Ault, Jesse~T}, \au{Warren, Patrick~B}, \au{Shin, Sangwoo} \&
  \au{Stone, Howard~A}} \yr{2017}  \at{Diffusiophoresis in one-dimensional
  solute gradients}.  \jt{Soft matter}  \bvol{13}~(47),  \pg{9015--9023}.

\bibitem[Barenblatt \& Isaakovich(1996)]{barenblatt1996scaling}
{\sc \au{Barenblatt, Grigory~Isaakovich} \& \au{Isaakovich,
  Barenblatt~Grigory}} \yr{1996} {\em Scaling, self-similarity, and
  intermediate asymptotics: dimensional analysis and intermediate
  asymptotics\/}.  \publ{Cambridge University Press}.

\bibitem[Bello {\em et~al.\/}(1994)Bello, Rezzonico \& Righetti]{bello1994use}
{\sc \au{Bello, Michael~S}, \au{Rezzonico, Roberta} \& \au{Righetti,
  Pier~Giorgio}} \yr{1994}  \at{Use of {Taylor-Aris} dispersion for measurement
  of a solute diffusion coefficient in thin capillaries}.  \jt{Science}
  \bvol{266}~(5186),  \pg{773--776}.

\bibitem[Ben-Yaakov(1972)]{ben1972diffusion}
{\sc \au{Ben-Yaakov, S}} \yr{1972}  \at{Diffusion of sea water ions—i.
  diffusion of sea water into a dilute solution}.  \jt{Geochimica et
  Cosmochimica Acta}  \bvol{36}~(12),  \pg{1395--1406}.

\bibitem[Bhattacharyya {\em et~al.\/}(2013)Bhattacharyya, Gopmandal, Baier \&
  Hardt]{bhattacharyya2013sample}
{\sc \au{Bhattacharyya, Somnath}, \au{Gopmandal, Partha~P}, \au{Baier, Tobias}
  \& \au{Hardt, Steffen}} \yr{2013}  \at{Sample dispersion in isotachophoresis
  with poiseuille counterflow}.  \jt{Physics of Fluids}  \bvol{25}~(2),
  \pg{022001}.

\bibitem[Biagioni {\em et~al.\/}(2022)Biagioni, Cerbelli \&
  Desmet]{biagioni2022shape}
{\sc \au{Biagioni, Valentina}, \au{Cerbelli, Stefano} \& \au{Desmet, Gert}}
  \yr{2022}  \at{Shape-enhanced open-channel hydrodynamic chromatography}.
  \jt{Analytical Chemistry}  \bvol{94}~(46),  \pg{15980--15986}.

\bibitem[Boudreau {\em et~al.\/}(2004)Boudreau, Meysman \&
  Middelburg]{boudreau2004multicomponent}
{\sc \au{Boudreau, Bernard~P}, \au{Meysman, Filip~JR} \& \au{Middelburg,
  Jack~J}} \yr{2004}  \at{Multicomponent ionic diffusion in porewaters:
  Coulombic effects revisited}.  \jt{Earth and Planetary Science Letters}
  \bvol{222}~(2),  \pg{653--666}.

\bibitem[Camassa {\em et~al.\/}(2021)Camassa, Ding, Kilic \&
  McLaughlin]{camassa2021persisting}
{\sc \au{Camassa, Roberto}, \au{Ding, Lingyun}, \au{Kilic, Zeliha} \&
  \au{McLaughlin, Richard~M}} \yr{2021}  \at{Persisting asymmetry in the
  probability distribution function for a random advection--diffusion equation
  in impermeable channels}.  \jt{Physica D: Nonlinear Phenomena}  \bvol{425},
  \pg{132930}.

\bibitem[Camassa {\em et~al.\/}(2010)Camassa, Lin \&
  McLaughlin]{camassa2010exact}
{\sc \au{Camassa, Roberto}, \au{Lin, Zhi} \& \au{McLaughlin, Richard~M}}
  \yr{2010}  \at{The exact evolution of the scalar variance in pipe and channel
  flow}.  \jt{Communications in Mathematical Sciences}  \bvol{8}~(2),
  \pg{601--626}.

\bibitem[Carney \& Engquist(2022)]{carney2022heterogeneous}
{\sc \au{Carney, Sean~P} \& \au{Engquist, Bj{\"o}rn}} \yr{2022}
  \at{Heterogeneous multiscale methods for rough-wall laminar viscous flow}.
  \jt{Communications in Mathematical Sciences}  \bvol{20}~(8),
  \pg{2069--2106}.

\bibitem[Casalini {\em et~al.\/}(2011)Casalini, Salvalaglio, Perale, Masi \&
  Cavallotti]{casalini2011diffusion}
{\sc \au{Casalini, Tommaso}, \au{Salvalaglio, Matteo}, \au{Perale, Giuseppe},
  \au{Masi, Maurizio} \& \au{Cavallotti, Carlo}} \yr{2011}  \at{Diffusion and
  aggregation of sodium fluorescein in aqueous solutions}.  \jt{The Journal of
  Physical Chemistry B}  \bvol{115}~(44),  \pg{12896--12904}.

\bibitem[Chatwin(1970)]{chatwin1970approach}
{\sc \au{Chatwin, PC}} \yr{1970}  \at{The approach to normality of the
  concentration distribution of a solute in a solvent flowing along a straight
  pipe}.  \jt{Journal of Fluid Mechanics}  \bvol{43}~(2),  \pg{321--352}.

\bibitem[Chatwin(1975)]{chatwin1975longitudinal}
{\sc \au{Chatwin, PC}} \yr{1975}  \at{On the longitudinal dispersion of passive
  contaminant in oscillatory flows in tubes}.  \jt{Journal of Fluid Mechanics}
  \bvol{71}~(3),  \pg{513--527}.

\bibitem[Cussler(2013)]{cussler2013multicomponent}
{\sc \au{Cussler, Edward~Lansing}} \yr{2013} {\em Multicomponent diffusion\/},
  ,  \vol{vol.~3}.  \publ{Elsevier}.

\bibitem[Deen(1998)]{deen1998analysis}
{\sc \au{Deen, William~Murray}} \yr{1998} {\em Analysis of transport
  phenomena\/}, ,  \vol{vol.~2}.  \publ{Oxford university press New York}.

\bibitem[Ding {\em et~al.\/}(2021)Ding, Hunt, McLaughlin \&
  Woodie]{ding2021enhanced}
{\sc \au{Ding, Lingyun}, \au{Hunt, Robert}, \au{McLaughlin, Richard~M} \&
  \au{Woodie, Hunter}} \yr{2021}  \at{Enhanced diffusivity and skewness of a
  diffusing tracer in the presence of an oscillating wall}.  \jt{Research in
  the Mathematical Sciences}  \bvol{8}~(3),  \pg{1--29}.

\bibitem[Ding \& McLaughlin(2022{\natexlab{{\em a\/}}})]{ding2022determinism}
{\sc \au{Ding, Lingyun} \& \au{McLaughlin, Richard~M}} \yr{2022{\natexlab{{\em
  a\/}}}}  \at{Determinism and invariant measures for diffusing passive scalars
  advected by unsteady random shear flows}.  \jt{Physical Review Fluids}
  \bvol{7}~(7),  \pg{074502}.

\bibitem[Ding \& McLaughlin(2022{\natexlab{{\em b\/}}})]{ding2022ergodicity}
{\sc \au{Ding, Lingyun} \& \au{McLaughlin, Richard~M}} \yr{2022{\natexlab{{\em
  b\/}}}}  \at{Ergodicity and invariant measures for a diffusing passive scalar
  advected by a random channel shear flow and the connection between the
  {Kraichnan}-{Majda} model and {Taylor}-{Aris} dispersion}.  \jt{Physica D:
  Nonlinear Phenomena}  \bvol{432},  \pg{133118}.

\bibitem[Ding \& McLaughlin(2023)]{ding2023dispersion}
{\sc \au{Ding, Lingyun} \& \au{McLaughlin, Richard~M.}} \yr{2023}
  \at{Dispersion induced by unsteady diffusion-driven flow in a parallel-plate
  channel}.  \jt{Phys. Rev. Fluids}  \bvol{8},  \pg{084501}.

\bibitem[Dutta \& Leighton(2001)]{dutta2001dispersion}
{\sc \au{Dutta, Debashis} \& \au{Leighton, David~T}} \yr{2001}  \at{Dispersion
  reduction in pressure-driven flow through microetched channels}.
  \jt{Analytical chemistry}  \bvol{73}~(3),  \pg{504--513}.

\bibitem[Eggers \& Fontelos(2008)]{eggers2008role}
{\sc \au{Eggers, Jens} \& \au{Fontelos, Marco~A}} \yr{2008}  \at{The role of
  self-similarity in singularities of partial differential equations}.
  \jt{Nonlinearity}  \bvol{22}~(1),  \pg{R1}.

\bibitem[Fischer(1969)]{fischer1969effect}
{\sc \au{Fischer, Hugo~B}} \yr{1969}  \at{The effect of bends on dispersion in
  streams}.  \jt{Water resources research}  \bvol{5}~(2),  \pg{496--506}.

\bibitem[GanOr {\em et~al.\/}(2015)GanOr, Rubin \&
  Bercovici]{ganor2015diffusion}
{\sc \au{GanOr, Nethanel}, \au{Rubin, Shimon} \& \au{Bercovici, Moran}}
  \yr{2015}  \at{Diffusion dependent focusing regimes in peak mode counterflow
  isotachophoresis}.  \jt{Physics of Fluids}  \bvol{27}~(7),  \pg{072003}.

\bibitem[Ghosal \& Chen(2010)]{ghosal2010nonlinear}
{\sc \au{Ghosal, Sandip} \& \au{Chen, Zhen}} \yr{2010}  \at{Nonlinear waves in
  capillary electrophoresis}.  \jt{Bulletin of mathematical biology}
  \bvol{72}~(8),  \pg{2047--2066}.

\bibitem[Ghosal \& Chen(2012)]{ghosal2012electromigration}
{\sc \au{Ghosal, S} \& \au{Chen, Z}} \yr{2012}  \at{Electromigration dispersion
  in a capillary in the presence of electro-osmotic flow}.  \jt{Journal of
  fluid mechanics}  \bvol{697},  \pg{436--454}.

\bibitem[Gopmandal \& Bhattacharyya(2015)]{gopmandal2015effects}
{\sc \au{Gopmandal, Partha~P} \& \au{Bhattacharyya, S}} \yr{2015}  \at{Effects
  of convection on isotachophoresis of electrolytes}.  \jt{Journal of fluids
  engineering}  \bvol{137}~(8).

\bibitem[Griffiths \& Stone(2012)]{griffiths2012axial}
{\sc \au{Griffiths, IM} \& \au{Stone, Howard~A}} \yr{2012}  \at{Axial
  dispersion via shear-enhanced diffusion in colloidal suspensions}.  \jt{EPL
  (Europhysics Letters)}  \bvol{97}~(5),  \pg{58005}.

\bibitem[Gupta {\em et~al.\/}(2019)Gupta, Shim, Issah, McKenzie \&
  Stone]{gupta2019diffusion}
{\sc \au{Gupta, Ankur}, \au{Shim, Suin}, \au{Issah, Luqman}, \au{McKenzie,
  Cameron} \& \au{Stone, Howard~A}} \yr{2019}  \at{Diffusion of multiple
  electrolytes cannot be treated independently: model predictions with
  experimental validation}.  \jt{Soft Matter}  \bvol{15}~(48),
  \pg{9965--9973}.

\bibitem[Hashemi {\em et~al.\/}(2018)Hashemi, Bukosky, Rader, Ristenpart \&
  Miller]{hashemi2018oscillating}
{\sc \au{Hashemi, Aref}, \au{Bukosky, Scott~C}, \au{Rader, Sean~P},
  \au{Ristenpart, William~D} \& \au{Miller, Gregory~H}} \yr{2018}
  \at{Oscillating electric fields in liquids create a long-range steady field}.
   \jt{Physical review letters}  \bvol{121}~(18),  \pg{185504}.

\bibitem[Hosokawa {\em et~al.\/}(2011)Hosokawa, Yamada, Johannesson \&
  Nilsson]{hosokawa2011development}
{\sc \au{Hosokawa, Yoshifumi}, \au{Yamada, Kazuo}, \au{Johannesson, Bj{\"o}rn}
  \& \au{Nilsson, Lars-Olof}} \yr{2011}  \at{Development of a multi-species
  mass transport model for concrete with account to thermodynamic phase
  equilibriums}.  \jt{Materials and Structures}  \bvol{44},  \pg{1577--1592}.

\bibitem[Ignatova \& Shu(2021)]{ignatova2021global}
{\sc \au{Ignatova, Mihaela} \& \au{Shu, Jingyang}} \yr{2021}  \at{Global
  solutions of the nernst--planck--euler equations}.  \jt{SIAM Journal on
  Mathematical Analysis}  \bvol{53}~(5),  \pg{5507--5547}.

\bibitem[Leaist(2017)]{leaist2017quinary}
{\sc \au{Leaist, Derek~G}} \yr{2017}  \at{Quinary mutual diffusion coefficients
  of aqueous mannitol+ glycine+ urea+ kcl and aqueous tetrabutylammonium
  chloride+ licl+ kcl+ hcl solutions measured by taylor dispersion}.
  \jt{Journal of Solution Chemistry}  \bvol{46}~(4),  \pg{798--814}.

\bibitem[Leaist \& Hao(1993)]{leaist1993diffusion}
{\sc \au{Leaist, Derek~G} \& \au{Hao, Ling}} \yr{1993}  \at{Diffusion in
  buffered protein solutions: combined nernst--planck and multicomponent fick
  equations}.  \jt{Journal of the Chemical Society, Faraday Transactions}
  \bvol{89}~(15),  \pg{2775--2782}.

\bibitem[Leaist \& MacEwan(2001)]{leaist2001coupled}
{\sc \au{Leaist, Derek~G} \& \au{MacEwan, Kimberley}} \yr{2001}  \at{Coupled
  diffusion of mixed ionic micelles in aqueous sodium dodecyl sulfate+ sodium
  octanoate solutions}.  \jt{The Journal of Physical Chemistry B}
  \bvol{105}~(3),  \pg{690--695}.

\bibitem[Lee {\em et~al.\/}(2021)Lee, Luner, Marzuola \&
  Harris]{lee2021dispersion}
{\sc \au{Lee, Garam}, \au{Luner, Alan}, \au{Marzuola, Jeremy} \& \au{Harris,
  Daniel~M}} \yr{2021}  \at{Dispersion control in pressure-driven flow through
  bowed rectangular microchannels}.  \jt{Microfluidics and Nanofluidics}
  \bvol{25}~(4),  \pg{1--11}.

\bibitem[Liu {\em et~al.\/}(2011)Liu, Shang \& Zachara]{liu2011multispecies}
{\sc \au{Liu, Chongxuan}, \au{Shang, Jianying} \& \au{Zachara, John~M}}
  \yr{2011}  \at{Multispecies diffusion models: A study of uranyl species
  diffusion}.  \jt{Water Resources Research}  \bvol{47}~(12).

\bibitem[Lyklema(2005)]{lyklema2005fundamentals}
{\sc \au{Lyklema, Johannes}} \yr{2005} {\em Fundamentals of interface and
  colloid science: soft colloids\/}, ,  \vol{vol.~5}.  \publ{Elsevier}.

\bibitem[Maex(2013)]{Maex2013Nernst}
{\sc \au{Maex, Reinoud}} \yr{2013} {\em Nernst-Planck Equation\/},  \pg{pp.
  1--7}.  \publ{New York, NY: Springer New York}.

\bibitem[Majda \& Kramer(1999)]{majda1999simplified}
{\sc \au{Majda, Andrew~J} \& \au{Kramer, Peter~R}} \yr{1999}  \at{Simplified
  models for turbulent diffusion: theory, numerical modelling, and physical
  phenomena}.  \jt{Physics reports}  \bvol{314},  \pg{237--574}.

\bibitem[Marbach \& Alim(2019)]{marbach2019active}
{\sc \au{Marbach, Sophie} \& \au{Alim, Karen}} \yr{2019}  \at{Active control of
  dispersion within a channel with flow and pulsating walls}.  \jt{Physical
  Review Fluids}  \bvol{4}~(11),  \pg{114202}.

\bibitem[Ngo-Cong {\em et~al.\/}(2015)Ngo-Cong, Mohammed, Strunin, Skvortsov,
  Mai-Duy \& Tran-Cong]{ngo2015higher}
{\sc \au{Ngo-Cong, D}, \au{Mohammed, FJ}, \au{Strunin, DV}, \au{Skvortsov, AT},
  \au{Mai-Duy, N} \& \au{Tran-Cong, T}} \yr{2015}  \at{Higher-order
  approximation of contaminant transport equation for turbulent channel flows
  based on centre manifolds and its numerical solution}.  \jt{Journal of
  Hydrology}  \bvol{525},  \pg{87--101}.

\bibitem[Oevreeide {\em et~al.\/}(2020)Oevreeide, Zoellner, Mielnik \&
  Stokke]{oevreeide2020curved}
{\sc \au{Oevreeide, Ingrid~H}, \au{Zoellner, Andreas}, \au{Mielnik, Michal~M}
  \& \au{Stokke, Bj{\o}rn~T}} \yr{2020}  \at{Curved passive mixing structures:
  a robust design to obtain efficient mixing and mass transfer in microfluidic
  channels}.  \jt{Journal of Micromechanics and Microengineering}
  \bvol{31}~(1),  \pg{015006}.

\bibitem[Poisson \& Papaud(1983)]{poisson1983diffusion}
{\sc \au{Poisson, A} \& \au{Papaud, A}} \yr{1983}  \at{Diffusion coefficients
  of major ions in seawater}.  \jt{Marine Chemistry}  \bvol{13}~(4),
  \pg{265--280}.

\bibitem[Price(1988)]{price1988theory}
{\sc \au{Price, William~E}} \yr{1988}  \at{Theory of the taylor dispersion
  technique for three-component-system diffusion measurements}.  \jt{Journal of
  the Chemical Society, Faraday Transactions 1: Physical Chemistry in Condensed
  Phases}  \bvol{84}~(7),  \pg{2431--2439}.

\bibitem[Ribeiro {\em et~al.\/}(2019)Ribeiro, Barros, Verissimo, Esteso \&
  Leaist]{ribeiro2019coupled}
{\sc \au{Ribeiro, Ana~CF}, \au{Barros, Marisa~CF}, \au{Verissimo, Luis~MP},
  \au{Esteso, Miguel~A} \& \au{Leaist, Derek~G}} \yr{2019}  \at{Coupled mutual
  diffusion in aqueous sodium (salicylate+ sodium chloride) solutions at 25°
  c}.  \jt{The Journal of Chemical Thermodynamics}  \bvol{138},  \pg{282--287}.

\bibitem[Rodrigo {\em et~al.\/}(2021)Rodrigo, Esteso, Ribeiro, Valente, Cabral,
  Verissimo, Musilova, Mracek \& Leaist]{rodrigo2021coupled}
{\sc \au{Rodrigo, M~Melia}, \au{Esteso, Miguel~A}, \au{Ribeiro, Ana~CF},
  \au{Valente, AJM}, \au{Cabral, Ana~MTDPV}, \au{Verissimo, Luis~MP},
  \au{Musilova, L}, \au{Mracek, A} \& \au{Leaist, Derek~G}} \yr{2021}
  \at{Coupled mutual diffusion in aqueous paracetamol+ sodium hydroxide
  solutions}.  \jt{Journal of Molecular Liquids}  \bvol{334},  \pg{116216}.

\bibitem[Rodrigo {\em et~al.\/}(2022)Rodrigo, Valente, Esteso, Cabral \&
  Ribeiro]{rodrigo2022ternary}
{\sc \au{Rodrigo, M~Melia}, \au{Valente, Artur~JM}, \au{Esteso, Miguel~A},
  \au{Cabral, Ana~MTDPV} \& \au{Ribeiro, Ana~CF}} \yr{2022}  \at{Ternary
  diffusion in aqueous sodium salicylate+ sodium dodecyl sulfate solutions}.
  \jt{The Journal of Chemical Thermodynamics}  \bvol{174},  \pg{106859}.

\bibitem[Schmuck \& Bazant(2015)]{schmuck2015homogenization}
{\sc \au{Schmuck, Markus} \& \au{Bazant, Martin~Z}} \yr{2015}
  \at{Homogenization of the poisson--nernst--planck equations for ion transport
  in charged porous media}.  \jt{SIAM Journal on Applied Mathematics}
  \bvol{75}~(3),  \pg{1369--1401}.

\bibitem[Sherman \& Morrison(1950)]{sherman1950adjustment}
{\sc \au{Sherman, Jack} \& \au{Morrison, Winifred~J}} \yr{1950}  \at{Adjustment
  of an inverse matrix corresponding to a change in one element of a given
  matrix}.  \jt{The Annals of Mathematical Statistics}  \bvol{21},
  \pg{124--127}.

\bibitem[Smith(1982)]{smith1982contaminant}
{\sc \au{Smith, Ronald}} \yr{1982}  \at{Contaminant dispersion in oscillatory
  flows}.  \jt{Journal of Fluid Mechanics}  \bvol{114},  \pg{379--398}.

\bibitem[Smith(1983)]{smith1983longitudinal}
{\sc \au{Smith, Ronald}} \yr{1983}  \at{Longitudinal dispersion coefficients
  for varying channels}.  \jt{Journal of Fluid Mechanics}  \bvol{130},
  \pg{299--314}.

\bibitem[Stone {\em et~al.\/}(2004)Stone, Stroock \&
  Ajdari]{stone2004engineering}
{\sc \au{Stone, Howard~A}, \au{Stroock, Abraham~D} \& \au{Ajdari, Armand}}
  \yr{2004}  \at{Engineering flows in small devices: microfluidics toward a
  lab-on-a-chip}.  \jt{Annu. Rev. Fluid Mech.}  \bvol{36},  \pg{381--411}.

\bibitem[Stroock {\em et~al.\/}(2002)Stroock, Dertinger, Ajdari, Mezic, Stone
  \& Whitesides]{stroock2002chaotic}
{\sc \au{Stroock, Abraham~D}, \au{Dertinger, Stephan~KW}, \au{Ajdari, Armand},
  \au{Mezic, Igor}, \au{Stone, Howard~A} \& \au{Whitesides, George~M}}
  \yr{2002}  \at{Chaotic mixer for microchannels}.  \jt{Science}
  \bvol{295}~(5555),  \pg{647--651}.

\bibitem[Tabrizinejadas {\em et~al.\/}(2021)Tabrizinejadas, Carrayrou,
  Saaltink, Baalousha \& Fahs]{tabrizinejadas2021validity}
{\sc \au{Tabrizinejadas, Sara}, \au{Carrayrou, Jerome}, \au{Saaltink,
  Maarten~W}, \au{Baalousha, Husam~Musa} \& \au{Fahs, Marwan}} \yr{2021}
  \at{On the validity of the null current assumption for modeling sorptive
  reactive transport and electro-diffusion in porous media}.  \jt{Water}
  \bvol{13}~(16),  \pg{2221}.

\bibitem[Taladriz-Blanco {\em et~al.\/}(2019)Taladriz-Blanco,
  Rothen-Rutishauser, Petri-Fink \& Balog]{taladriz2019precision}
{\sc \au{Taladriz-Blanco, Patricia}, \au{Rothen-Rutishauser, Barbara},
  \au{Petri-Fink, Alke} \& \au{Balog, Sandor}} \yr{2019}  \at{Precision of
  taylor dispersion}.  \jt{Analytical chemistry}  \bvol{91}~(15),
  \pg{9946--9951}.

\bibitem[Taylor(1953)]{taylor1953dispersion}
{\sc \au{Taylor, Geoffrey~Ingram}} \yr{1953}  \at{Dispersion of soluble matter
  in solvent flowing slowly through a tube}.  \jt{Proceedings of the Royal
  Society of London. Series A. Mathematical and Physical Sciences}
  \bvol{219}~(1137),  \pg{186--203}.

\bibitem[Taylor(2012)]{taylor2012random}
{\sc \au{Taylor, Michael}} \yr{2012}  \at{Random walks, random flows, and
  enhanced diffusivity in advection-diffusion equations}.  \jt{Discrete \&
  Continuous Dynamical Systems-B}  \bvol{17}~(4),  \pg{1261}.

\bibitem[Tournassat {\em et~al.\/}(2020)Tournassat, Steefel \&
  Gimmi]{tournassat2020solving}
{\sc \au{Tournassat, Christophe}, \au{Steefel, Carl~I} \& \au{Gimmi, Thomas}}
  \yr{2020}  \at{Solving the nernst-planck equation in heterogeneous porous
  media with finite volume methods: Averaging approaches at interfaces}.
  \jt{Water resources research}  \bvol{56}~(3),  \pg{e2019WR026832}.

\bibitem[Vanysek(1993)]{vanysek1993ionic}
{\sc \au{Vanysek, Petr}} \yr{1993}  \at{Ionic conductivity and diffusion at
  infinite dilution}.  \jt{CRC hand book of chemistry and physics}  \pg{pp.
  5--92}.

\bibitem[Vedel \& Bruus(2012)]{vedel2012transient}
{\sc \au{Vedel, S{\o}ren} \& \au{Bruus, Henrik}} \yr{2012}  \at{Transient
  {Taylor--Aris} dispersion for time-dependent flows in straight channels}.
  \jt{Journal of fluid mechanics}  \bvol{691},  \pg{95--122}.

\bibitem[Wang \& Roberts(2013)]{wang2013self}
{\sc \au{Wang, Wei} \& \au{Roberts, Anthony~J}} \yr{2013}  \at{Self-similarity
  and attraction in stochastic nonlinear reaction-diffusion systems}.  \jt{SIAM
  Journal on Applied Dynamical Systems}  \bvol{12}~(1),  \pg{450--486}.

\bibitem[Wu \& Chen(2014)]{wu2014approach}
{\sc \au{Wu, Zi} \& \au{Chen, GQ}} \yr{2014}  \at{Approach to transverse
  uniformity of concentration distribution of a solute in a solvent flowing
  along a straight pipe}.  \jt{Journal of Fluid Mechanics}  \bvol{740},
  \pg{196--213}.

\bibitem[Yotsukura \& Sayre(1976)]{yotsukura1976transverse}
{\sc \au{Yotsukura, Nobuhiro} \& \au{Sayre, William~W}} \yr{1976}
  \at{Transverse mixing in natural channels}.  \jt{Water Resources Research}
  \bvol{12}~(4),  \pg{695--704}.

\bibitem[Young \& Jones(1991)]{young1991shear}
{\sc \au{Young, WR~a} \& \au{Jones, Scott}} \yr{1991}  \at{Shear dispersion}.
  \jt{Physics of Fluids A: Fluid Dynamics}  \bvol{3}~(5),  \pg{1087--1101}.

\bibitem[Yuan-Hui \& Gregory(1974)]{yuan1974diffusion}
{\sc \au{Yuan-Hui, Li} \& \au{Gregory, Sandra}} \yr{1974}  \at{Diffusion of
  ions in sea water and in deep-sea sediments}.  \jt{Geochimica et cosmochimica
  acta}  \bvol{38}~(5),  \pg{703--714}.

\end{thebibliography}

\end{document}